\documentclass[fleqn,usenatbib]{mnras}
\usepackage[T1]{fontenc}
\usepackage{ae,aecompl}
\usepackage{graphicx}	% Including figure files
\usepackage{amsmath}	% Advanced maths commands
\usepackage{amssymb}	% Extra maths symbols
\usepackage{soul}

\title[The assembly bias of ELGs]{The assembly bias of emission line galaxies}

\author[Esteban Jim\'{e}nez et al.]{
Esteban Jim\'{e}nez $^{1,2}$\thanks{E-mail: esteban.jimenez@icrar.org},
Nelson Padilla$^{1,3}$,
Sergio Contreras$^{4}$,
Idit Zehavi$^{5}$,
\newauthor
Carlton M. Baugh$^{6}$ and
\'{A}lvaro Orsi $^{7,8}$
\\
$^{1}$Instituto de Astrof\'{i}sica, Pontificia Universidad Cat\'{o}lica de Chile, Santiago, Chile\\
$^{2}$International Centre for Radio Astronomy Research (ICRAR), University of Western Australia, Crawley, WA 6009, Australia\\
$^{3}$Centro de Astro-Ingenier\'{i}a, Pontificia Universidad Cat\'{o}lica de Chile, Santiago, Chile\\
$^{4}$Donostia International Physics Center (DIPC), Manuel Lardizabal pasealekua 4, 20018 Donostia, Basque Country, Spain\\
$^{5}$Department of Physics, Case Western Reserve University, Cleveland, OH 44106, USA\\
$^{6}$Institute for Computational Cosmology, Department of Physics, Durham University, South Road, Durham, DH1 3LE, UK\\
$^{7}$Centro de Estudios de F\'{i}sica del Cosmos de Arag\'{o}n. Plaza San Juan 1, planta 2, 44001 Teruel, Spain\\
$^{8}$PlantTech Research Institute Limited. South British House, 4th Floor, 35 Grey Street, Tauranga 3110, New Zealand\\
}

\date{Accepted XXX. Received YYY; in original form ZZZ}

\pubyear{2020}

\begin{document}
\label{firstpage}
\pagerange{\pageref{firstpage}--\pageref{lastpage}}
\maketitle

% Abstract of the paper
\begin{abstract}
The next generation of spectroscopic surveys will target emission-line galaxies (ELGs) to produce constraints on cosmological parameters. We study the large scale structure traced by ELGs using a combination of a semi-analytical model of galaxy formation, a code that computes the nebular emission from HII regions using the  properties of the interstellar medium, and a large-volume, high-resolution N-body simulation. We consider fixed number density samples where galaxies are selected by either their H$\alpha$, [OIII]$\lambda 5007$ or [OII]$\lambda \lambda 3727-3729$ emission line luminosities. We investigate the assembly bias signatures of these samples, and compare them to those of stellar mass and SFR selected samples. Interestingly, we find that the [OIII]- and [OII]-selected samples display scale-dependent bias on large scales and that their assembly bias signatures are also scale-dependent. Both these effects are more pronounced for lower number density samples. The [OIII] and [OII] emitters that contribute most to the scale dependence tend to have a low gas-phase metallicity and are preferentially found in low-density regions. We also measure the baryon acoustic oscillation (BAO) feature and the $\beta$ parameter related to the growth rate of overdensities. We find a slight tendency for the BAO peak to shift toward smaller scales for [OII] emitters and that $\beta$ is scale-dependent at large scales. Our results suggest that ELG samples include environmental effects that should be modelled in order to remove potential systematic errors that could affect the estimation of cosmological parameters.
\end{abstract}


%%

%%%%%%%%%%%%%%%%% BODY OF PAPER %%%%%%%%%%%%%%%%%%

\section{Introduction} 

Mapping the Universe using photometric and spectroscopic surveys allows us to measure the cosmic large-scale structure which encodes valuable cosmological information. However, since the galaxy distribution is not a direct tracer of the underlying density field it is essential to understand the connection between galaxies and dark matter haloes to obtain an accurate interpretation of the Universe (for a review see \citealt{Wechsler18}). A relevant statistical property of the galaxy distribution is the clustering signal which is often quantified using the two-point correlation function (2PCF). Measuring galaxy clustering allows us to extract two pieces of information that can be used to constrain the cosmological model \citep{Weinberg13}: (i) the scale of standard ruler features, such as the baryonic acoustic oscillation (BAO) peak, from which we can obtain the cosmic expansion history, and (ii) the magnitude of redshift-space distortions (RSD) in the clustering signal, which are driven by the rate at which structure grows. Both of these quantities depend on the amount of dark matter and dark energy.  

Precise measurements of BAO and RSD are difficult to obtain because cosmic (sample) variance and shot noise are significant when the sampled volume is small \citep{Kaiser86a}. The Sloan Digital Sky Survey (SDSS) and the 2dF Galaxy Redshift Survey (2dFGRS) were the first to observe hundreds of thousands of galaxies in large volumes, obtaining convincing detections of the BAO \citep{Eisenstein05, Cole05}. Observations have continued throughout the last fifteen years mostly using massive luminous red galaxies (LRGs) as the tracers of the large-scale structure \citep[e.g][]{Eisenstein11, Zehavi11, Dawson13, Bautista18}. 

Advances in wide field spectroscopy have opened up the opportunity to trace the large-scale structure using emission-line galaxies (ELGs). The nebular emission of these galaxies is produced by gas in HII regions which is photoionized by radiation from young stars. Some  emission line luminosities have therefore been used to infer star formation rates, although in general the relation between line emission and SFR can be complicated as the emission depends on the local properties of the ISM such as gas metallicity, temperature and density \citep[e.g][]{Levesque10, Gutkin:2016, Byler:2017} and on the attenuation by dust of the line luminosity. Moreover, ELGs do not trace the field in the same way as LRGs; ELGs tend to reside in low mass haloes \citep{Favole16, gp18} and live in filaments and sheets rather than in the knots of the cosmic web occupied by LRGs \citep{GP20}.  

Typically, given the depth of upcoming surveys with spectrographs that operate in the optical, ELG catalogues have redshift distributions that peak around $z \sim 1$ which, in combination with Ly$\alpha$ emitters at $z \gtrsim 2$, enables us to investigate the history of cosmic expansion at previously unexplored epochs. The SDSS-IV/eBOSS survey \citep{Dawson16} provides one of the largest ELG catalogues to date, and the next generation of surveys like DESI \citep{DESI16} and Euclid \citep{Laureijs:2011} will detect millions of ELGs. This will potentially enable us to measure cosmological parameters with exceptional precision.   

To fully exploit the future ELG data it is essential to understand any systematic effects that may influence the inferred cosmological constraints. Galaxy formation models can be used for this purpose, to provide insights into the galaxy-halo connection with the by-product of testing different prescriptions for the physical processes that regulate galaxy evolution. A useful approach to explore galaxy formation is to use semi-analytical models \citep[SAMs][]{cole2000, Baugh06-rv, Somervile08}. These models use simplified descriptions of physical processes that shape the fate of baryons within the dark matter halo merger trees extracted from N-body dark matter only simulations, expressed in a set of coupled differential equations with parameters to encapsulate ``sub-grid'' physics. SAMs can successfully reproduce, amongst other things, the observed luminosity and stellar mass functions \citep[e.g][]{Croton06, Henriques15, Croton16, Stevens16, Lagos18, Baugh19}. Alternatively, hydrodynamical simulations offer a complementary approach to follow baryonic physics, which in general requires fewer assumptions and approximations than are needed in SAMs, but which nevertheless still appeal to using sub-grid recipes for unresolved processes \citep[e.g][]{Vogelsberger14, Schaye15, Nelson18}. Due to the higher computational overhead of hydrodynamical simulations compared with SAMs, the largest volumes probed by state-of-the-art hydrodynamical simulations are still 10-100 times smaller than the typical SAM volume. 

Another approach used to connect galaxies with their host haloes is to employ an empirical model such as the halo occupation distribution (HOD) framework \citep[e.g][]{Benson00, Scoccimarro01, Berlind02, Kravtsov04, Zheng05}. The HOD provides an empirical relation between the average number of galaxies $\rm N$ hosted by haloes with mass $\rm M$. This relation is characterized by a probability distribution $\rm P(N|M)$ that depends on the redshift, number density and selection criteria of a galaxy sample. Here, the standard assumption is that the galaxy content depends only on halo mass, but this may not be true if the galaxy distribution correlates with the assembly history of haloes. N-body simulations have shown that the clustering of dark matter haloes does depend on secondary halo properties like formation time, concentration and spin \citep[e.g][]{Gao05, Gao07, Wechsler06}, an effect called halo assembly bias. Likewise, the manifestation of assembly bias in galaxy clustering, commonly referred to as galaxy assembly bias, has been found both in SAMs \citep[e.g][]{Croton07, contreras19,  Zehavi18, Zehavi19, Jimenez19, Xu20b} and hydrodynamical simulations \citep[e.g][]{Artale18, Xu20a, Montero20}.  Assessing the existence of assembly bias in the real Universe is an important task; cosmological constraints from future surveys will most likely be limited by how well we can model the observations rather than the precision of the measurements.    

In general, assembly bias enhances the clustering amplitude on large scales for stellar mass selected samples and suppresses it for  SFR selections \citep{contreras19, Contreras20}.  However, as we found in these studies, these trends can change depending on the number density and redshift of the sample. So far, there are no direct measurements of the assembly bias signature in ELG catalogues, but, in principle, the effect should be similar to that reported for  SFR selections as ELGs are a subset of star-forming galaxies.  

Here we aim to study the large-scale structure of ELGs by measuring the clustering and galaxy assembly bias signature of these samples. We employ galaxies from the SAG semi-analytical model \citep{Cora18} run on the MultiDark Planck cosmological simulation \citep{Klypin16}. The total simulated volume is ${\rm (1 }\ h^{-1} {\rm Gpc)^3}$, so the effect of the sample variance is greatly reduced. Thus, we can accurately sample the 2PCF up to scales of the BAO feature and determine whether or not the impact of assembly bias from ELG selections is significant. We calculate the nebular emission in SAG galaxies using the {\tt GET\_EMLINES} code from \citet{Orsi14}, and then store the H$\alpha$, [OIII] and [OII] line emission luminosities. These emission lines at $\rm z \sim 1$ correspond either to the near-infrared and optical range sampled by DESI and Euclid, respectively.   

The layout of the paper is as follows. Section 2 describes the SAG galaxy formation model and the N-body simulation in which it is implemented, along with how we define our galaxy samples, while in Section 3 we compare and characterize these samples. In Section 4 we show the assembly bias signatures, and in Section 5 we study a possible origin for features in the assembly bias of ELGs. The BAO and the $\beta$ parameter, which quantifies the strength of anisotropies produced by the RSD, are shown in Section 6 for each sample. We conclude in Section 7. Brief discussions about results from other SAMs, and ELG sample completeness, are presented in Appendix A and B, respectively.   

Throughout the paper masses are measured in $h^{-1} \rm M_{\odot}$, the SFR is measured in $ h^{-1} \rm M_{\odot}yr^{-1}$ and distances are measured in $h^{-1} \rm Mpc$ and are in comoving units.   

%%%%%%%%%%%%%%%%%%%%%%%%%%%%%%%%%%%%%%%%%%%%%%
\section{Simulation data}

\subsection{The galaxy formation model: SAG}

Here we use the Semi-Analytical Galaxies (SAG) model of galaxy formation \citep{Cora06}. Semi-analytical models use the merger trees extracted from N-Body simulations to model the main physical processes involved in the evolution of a galaxy, such as the treatment of radiative cooling of hot gas, star formation, feedback effects from supernovae and active galactic nucleus (AGN) feedback, chemical enrichment of the gas, the growth of supermassive black holes, and the impact from galaxy mergers, among others. 

The version of SAG used in this work is the one presented in \citet{Cora18}, which uses the outputs of the MultiDark2 cosmological simulation (hereafter MDPL2, \citealt{Klypin16}, see section 2.2 for more details). The main output of the simulation and the SAM are publicly available\footnote{\url{http://www.cosmosim.org}} as a part of the MultiDark comparison project \citep{knebe18}. The SAG SAM was originally presented in \cite{Cora06} and is based on the model by \cite{Springel01}.  Since then the code has been through several updates \citep{Lagos08, Tecce10, Orsi14, Padilla14, Munoz15, Gargiulo2015} and is capable of reproducing observations both at low and high redshift. One of the key features of this model is the use of particle swarm optimization technique (PSO) to automatically set  the model parameters by requiring the output to fit several observables \citep{Ruiz15}.

The galaxy properties produced by this model include stellar mass, cold gas, black hole and bulge masses, the average and instantaneous star formation rates (SFR), where the latter corresponds to the SFR in the most recent time sub-step, which is a subdivision of the timestep between the simulation snapshots.  Sub-step sizes range from 5 and 15 Myr, whereas the time between snapshots is of the order of 100 Myr. These quantities are computed separately for disks and bulges, where the former are the result of quiescent star formation in cooled gas disks and the latter form via starburst episodes. The gas-phase metallicity for both components is calculated by modelling the chemical enrichment of the ISM, which takes into account mass loss from massive stars and supernovae. These two ways to compute the SFR are important when computing the emission line fluxes of the galaxies (see section 2.3 for more details).

\subsection{The MultiDark Planck 2 simulation}
\label{subsec: MDPL2}

As mentioned in the previous section, SAG was run on the halo merger trees from the {\tt MULTIDARK} simulation MDPL2 \citep{Klypin16}. The MDPL2 adopts a $\Lambda$CDM Universe, charaterized by Planck cosmological parameters \citep{Plank14}: $\Omega_{\rm m}=0.307$, $\Omega_{\rm b} = 0.048$, $\Omega_{\Lambda} = 0.693$, $h=0.678$ and $n_{\rm s}=0.96$. The simulation follows $3840^3$ particles within a cubic box of comoving side-length $ 1\ h^{-1}\rm Gpc$, with a mass resolution of ${\rm m_p = 1.51\times 10^{9}} h^{-1} {\rm M_{\odot}}$. The particles are followed from $z=120$ until z=0 and their positions and velocities are output at 126 snaphots. The dark matter haloes are identified with the {\tt ROCKSTAR} halo finder \citep{Behroozi2013a}, and the {\tt CONSISTENT TREES} code \citep{Behroozi2013b} is used to construct the merger trees. These halo finder and halo merger tree algorithms identify objects in phase space, keeping a better track of the substructures after their infall.

The large cosmological volume of the MDPL2 allows us to make accurate clustering measurements up to scale separations that encapsulate useful cosmological information.

\subsection{The calculation of nebular emission}

To model the nebular emission of the star-forming galaxies we use the {\tt GET\_EMLINES} model\footnote{https://github.com/aaorsi/get\_emlines} introduced by \citet{Orsi14} (hereafter O14) to post-process the output from the SAG model. The {\tt GET\_EMLINES} code uses the output of the photoionisation code {\tt MAPPINGS-III} \citep{Dopita95, Groves04}, as tabulated by \cite{Levesque10}. {\tt MAPPINGS-III} predicts the nebular emission from HII regions. The grid calculated by \cite{Levesque10} tabulates the emission line fluxes as a function of the gas-phase metallicity and the ionization parameter of the HII region, $q$. O14 uses the gas metallicities of the bulges and disks from the SAG galaxies, but it does not predict the ionization 
parameter within individual HII regions due to a lack of resolution to resolve the internal structure of galaxies. Instead, O14 advocated a model in which $q$ could be inferred from the gas-phase metallicity, an assumption which is inspired by observational results which suggest that $q$ is anti-correlated with the metallicity of the cold star-forming gas \citep[e.g][]{Nagao06, Groves10, Shim13}.
O14 show that parameters selected in their model to calculate $q$ from the gas phase metallicity allowed the SAG model to reproduce the locus of star forming galaxies in the so-called BPT diagram relating the line ratios [OIII$\lambda$5007]/H$\beta$ and [NII$\lambda$6854]/H$\alpha$ \citep{BPT:1981}.cNote that O14 showed that the predictions were robust to substantial perturbations to the parameter values in the model for $q$. 
Ideally, the {\tt GET\_EMLINES} code uses the instantaneous SFR rather than a time-averaged SFR, 
as the instantaneous SFR is a better indicator of the number of Lyman continuum photons which make up the ionising radiation field and, as a consequence, of the H$\alpha$ luminosity. Nevertheless, \citet{Favole20} used the averaged SFR predicted by the {\tt SAGE} \citep{Croton16} and {\tt GALACTICUS} \citep{Benson12} SAMs to infer the [OII] line emission, and found reasonable agreement with observational data at $z\sim 1$.

We use {\tt GET\_EMLINES} and instantaneous SFRs to compute the luminosities for the H$\alpha$, [OIII]$\lambda 5007$, [OII]$\lambda \lambda 3727-3729$ and [NII]$\lambda \lambda 6548-6584$ emission lines (hereafter H$\alpha$, [OIII], [OII] and [NII] respectively). As these values are calculated separately for the bulge and disk components of a galaxy, we sum these contributions to obtain the total nebular emission for each galaxy. Note that we do not apply any attenuation to the line luminosity. 

\subsection{The galaxy samples}

We use galaxy samples characterised by fixed number densities of objects. We achieve this by ranking the galaxies from the highest to lowest values of a given property (e.g. their emission line luminosity, averaged SFR or stellar mass), and then retaing only those galaxies above a threshold value that corresponds to the desired number density. The number densities used in this work are $n=0.001\ h^3{\rm Mpc^{-3}}$, $ n=0.00316\ h^3 {\rm Mpc^{-3}}$ and $n=0.01\ h^3 {\rm Mpc^{-3}}$.

\begin{figure*}
    \centering
    \includegraphics[width=0.98\textwidth]{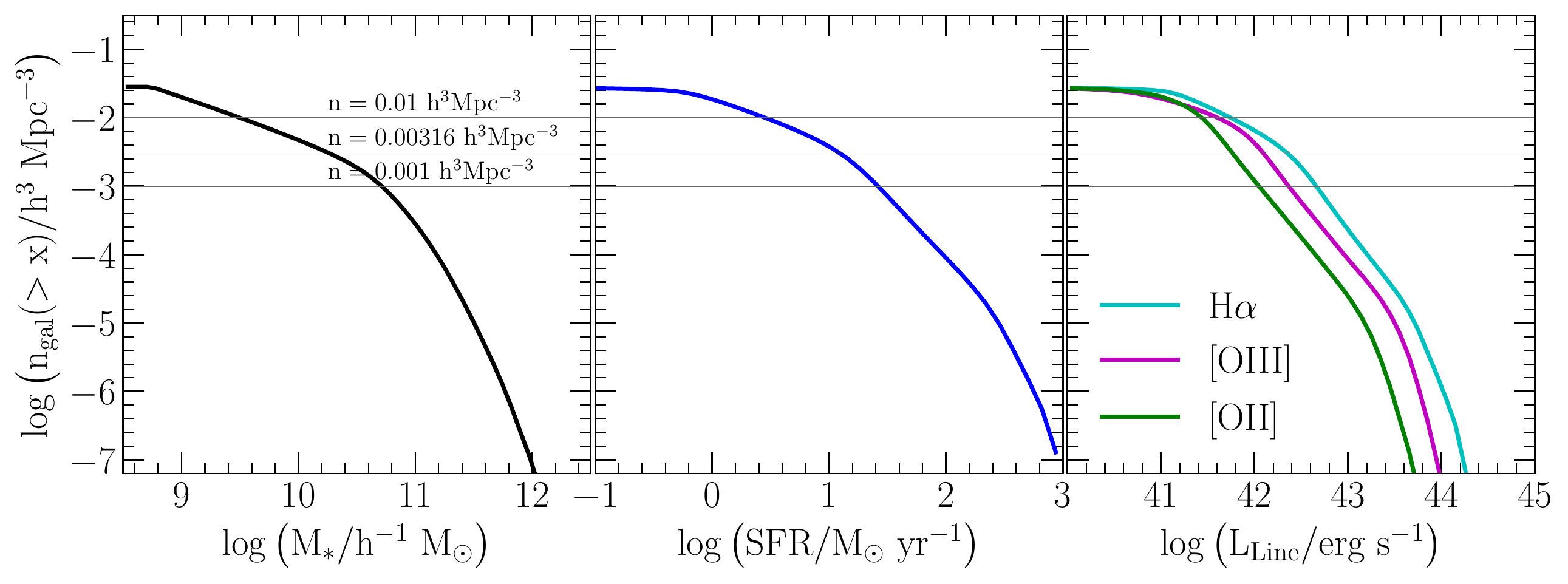}
    \caption{The cumulative distribution functions of SAG galaxies selected by stellar mass ({\it left}), SFR ({\it middle}) and $\rm H\alpha$, [OIII], [OII] line luminosities ({\it right}). The horizontal dashed lines indicate the number densities used to define our galaxy samples. Note the plateau at low SFR and $\rm L_{Line}$, which is due to a stellar mass cut in the parent catalogue, which is discussed further in Appendix B.}
    \label{fig: CumFunctions}
\end{figure*}

As the samples selected by line emission are star-forming galaxies, it is expected that they will have some overlap with galaxies selected by their SFR. Hence, we also include SFR selected samples and perform the same analysis on these as carried out for the ELG samples. We also consider stellar-mass selected samples to allow further comparisons with assembly bias signatures which have been studied in several galaxy formation models \citep[e.g][]{Artale18, Zehavi18, contreras19}. The cumulative distribution functions for the different selections are shown in Fig.~\ref{fig: CumFunctions}. The number densities adopted to define our samples are shown by the horizontal thin dashed lines.  Note that the galaxies included in each sample are those to the right of the intersection between the cumulative functions with the horizontal lines. We also include a selection based on the combined luminosity of the H$\alpha$+doublet [NII]$\lambda \lambda 6548-6584$ lines; this combination mimics the flux that the Euclid mission will capture for its sources due to its limited spectral resolution. This allows us to assess the contribution of the [NII] flux to the predictions for H$\alpha$ (see also \citealt{Merson:2018}).

The data used here correspond to a subsample of SAG\footnote{ http://skiesanduniverses.org/} where galaxies were  selected with a stellar-mass cut of ${\rm 5\times 10^8}\ h^{-1}{\rm M_{\odot}}$. 
This cut affects the completeness of SFR and ELG selections. In Appendix B we explain that this has a negligible impact on the trends and results we obtain.
%\bf Do we need this paragraph here? Distraction?}  {\color{red} \bf EJ: I think we should include this paragraph because Fig. 1 shows the completeness issue at low SFRs. Then, if questions related to this arise while looking at the figure (or reading this subsection), readers can look for the answers in Appendix B.}  

\section{Properties of ELG selected samples}

\begin{figure}
    \centering
    \includegraphics[width=\columnwidth]{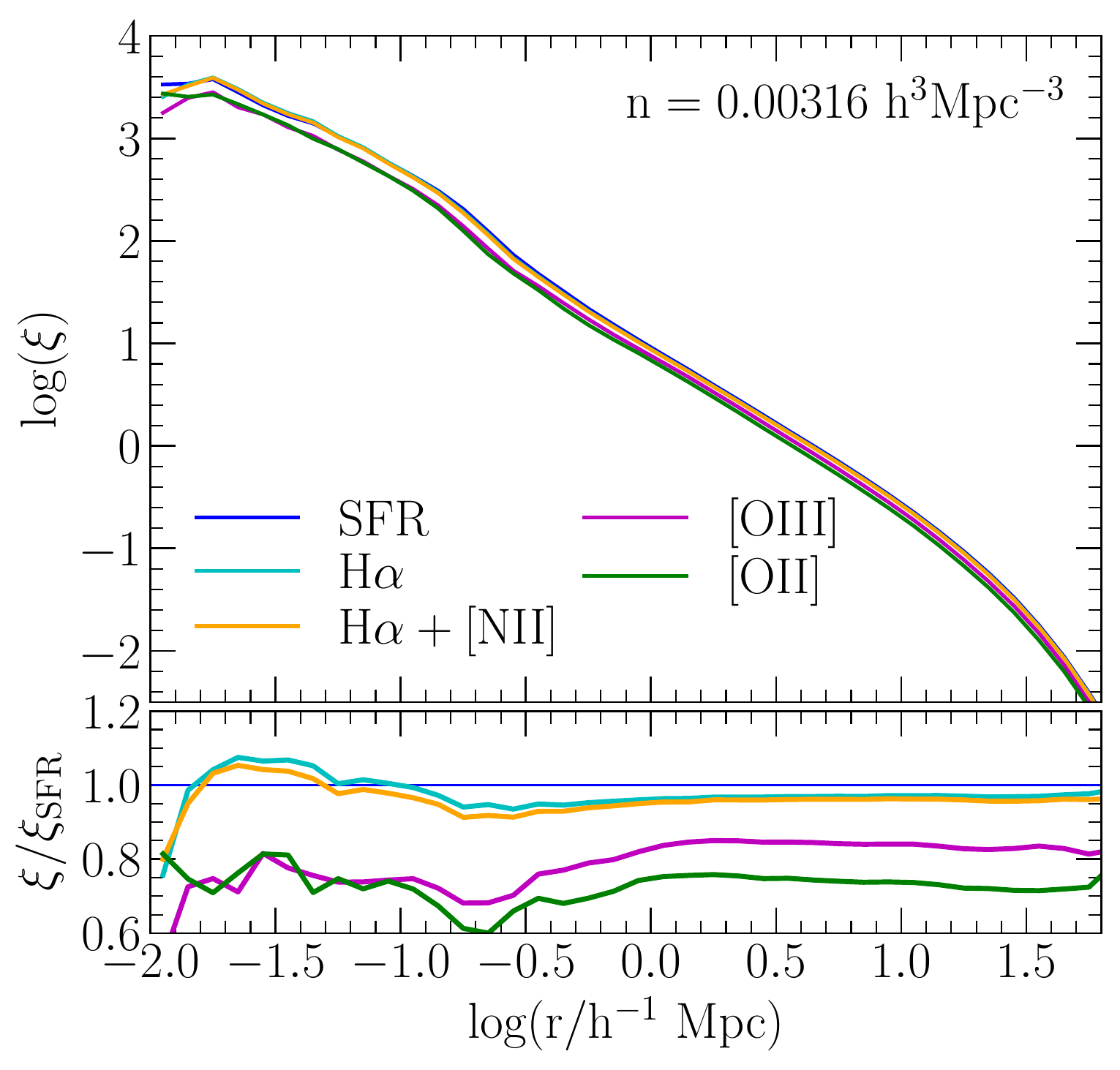}
    \caption{The 2PCFs for SAG samples selected according to different properties, as indicated by the key; in each case the number density is $n = 0.00316\ h^3{\rm Mpc^{-3}}$. The bottom panel shows the ratio of the correlation functions measured for  each sample with respect to the SFR-selected sample.}
    \label{fig: xi_SAG}
\end{figure}

To assess the level of similarity between the ELG and star-forming galaxy samples we compare their two-point correlation functions (2PCF).
The 2PCF measures the excess probability of finding a pair of objects at a separation $r$ compared to a homogeneously distributed sample. We measure the 2PCF using the {\tt CorrFunc} public code presented in \citet{Sinha17}\footnote{\url{https://corrfunc.readthedocs.io/en/master/}}. The resulting 2PCF are shown in the main panel of Fig.~\ref{fig: xi_SAG} for the SFR and ELG samples with number density $n=0.00316\ h^3 {\rm Mpc^{-3}}$; the sub-panel shows the ratios between these measurements and the 2PCF of the SFR selected sample. The first impression is that the shapes of the 2PCFs are largely similar, irrespective of separation, with variations within $10\%$. On large scales, the differences are mostly due to the different bias parameters of the samples, with the [OIII] and [OII] selected samples showing weaker clustering than the H$\alpha$ and SFR samples. On small scales, the differences may be due to the additional dependence on the physical conditions in the ISM; the H$\alpha$  emission mostly traces the SFR, but the [OIII] and [OII] emission also depends on the cold gas metallicity. Hence, differences in the one-halo terms suggest a possible connection between the spatial distribution of ELGs and the properties of their ISM. As can be seen from Fig.~\ref{fig: xi_SAG}, the difference in the 2PCF compared with that measured for the SFR selected sample depends on which line is chosen to construct the sample. Both the H$\alpha$ and H$\alpha$ + [NII] selections result in an amplitude for the 2PCF that is similar to that found  for the SFR selection. This is expected as the strength of H$\alpha$ emission is almost an instantaneous measure of the star formation rate, with little dependence on the metallicity of the star forming gas.  The 2PCFs measured for the [OIII] and [OII] selections show a bigger difference from that found for the star formation sample, with an amplitude reduction of $\sim 20$ and $30$ per cent, respectively. This change in the effective bias parameter of these samples is related to the selection of galaxies with specific combinations of SFR and gas metallicity, as we demonstrate below.

One way to interpret the 2PCF is by using the halo occupation distribution (HOD) framework. The HOD characterises a galaxy population via the halo occupation function, the average number of galaxies as a function of the host halo mass. Whereas HODs are typically used as an empirical model with parameters which are set to reproduce the measured abundance and 2PCF of a galaxy samples, SAMs predict the form of the HOD. So, in the case of SAMs, the HOD formalism produces a concise description of the model output that can be readily interpreted in relation to the 2PCF. In general, this function is separated into the contribution from central galaxies and from satellites with specific forms that depend on the selection criteria \citep{Zheng05}. For example, when samples are defined by luminosity or stellar mass cuts, the HOD for central galaxies follows a step-like form. When samples are defined by SFR or colour cuts, on the other hand, the HOD of centrals reaches a peak followed by a dip to values below unity as the halo mass increases \citep[e.g][]{Zehavi11, contreras13, gp18, Jimenez19}. The output of a SAM for these selections can be tabulated as an HOD, without having to adopt a particular parametric form, which is very powerful when considering selections for which there is little available data, such as ELGs.  

\begin{figure}
    \centering
    \includegraphics[width=\columnwidth]{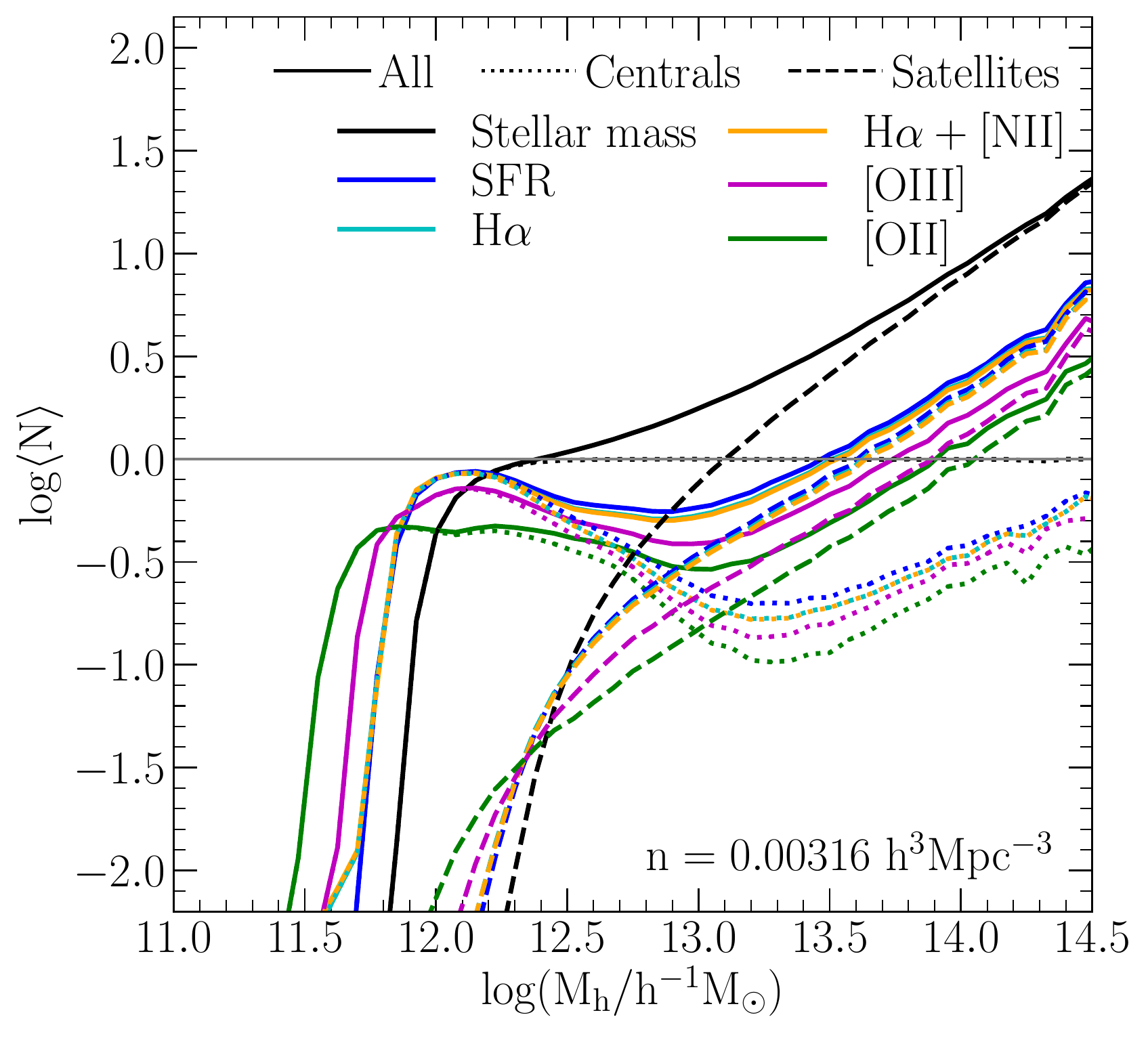}
    \caption{The HODs predicted by the SAG model for all galaxies (solid), centrals (dotted) and satellites (dashed), with different colours indicating  different galaxy selections, as shown  by the figure key. All samples have a number density of $n=0.00316\ h^3 {\rm Mpc^{-3}}$.}
    \label{fig: HODs}
\end{figure}

Fig.~\ref{fig: HODs} shows the HODs predicted by the SAG model for SFR and ELG selections with a number density $n=0.00316\ h^3{\rm Mpc^{-3}}$. We also show results for a stellar mass selected sample to illustrate the differences compared to the star-forming and ELG samples. The stellar mass selected sample in the figure shows the canonical step-like form for the HODs of centrals, with the HOD for satellites exhibiting a power-law behaviour. In contrast, the ELG selections show a peak in the HOD of central galaxies, which shifts to lower masses for the [OIII] and [OII] selections. This indicates that model ELGs are mostly hosted by low halo masses, consistent with previous results from simulations \citep[e.g][]{gp18}. For large halo masses, the HODs of both centrals and satellites increase with halo mass. 

\begin{figure}
    \centering
    \includegraphics[width=\columnwidth]{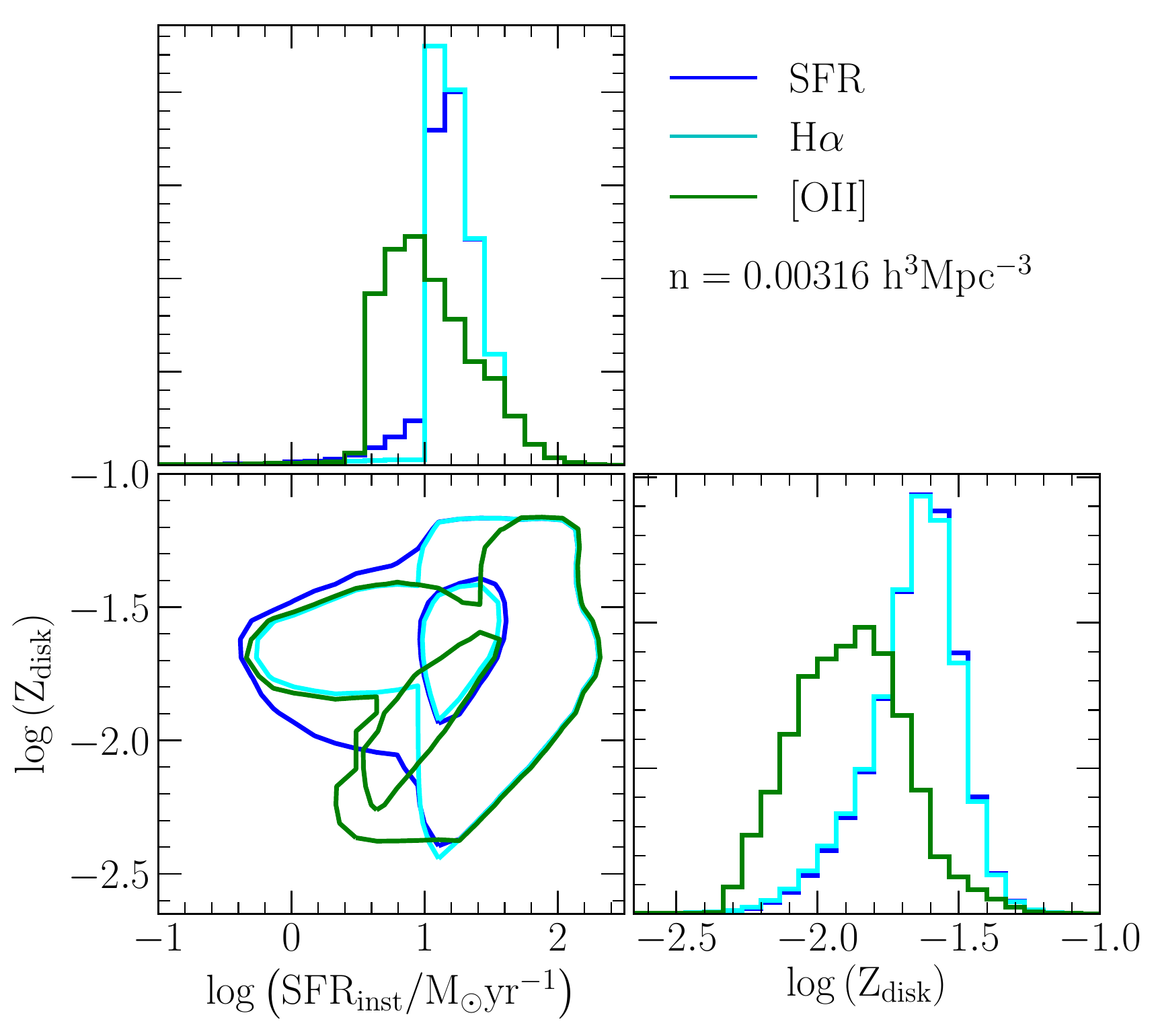}
    \caption{Distributions of instantaneous SFR ($\rm SFR_{inst}$) and cold gas metallicity of the disk ($\rm Z_{disk}$)  for SAG samples with fixed number density $n= 0.00316\ h^3 {\rm Mpc^{-3}}$. Different colours correspond to different selections as shown by the key; the density contours enclose 68\% and 99\% of the distribution of galaxies. The corner panel shows the bivariate distributions, whereas the top and right panels show the marginalised distributions of instantaneous SFR and cold gas metallicity, respectively.}
    \label{fig: corner-plot}
\end{figure}

The overlap between the SFR and ELG-selected samples can be quantified by analysing the similarities in their galaxy properties. Fig.~\ref{fig: corner-plot} shows the distribution of galaxies in the instantaneous SFR vs. gas metallicity plane for the SFR, H$\alpha$ and [OII] selections, in all cases with a number density $n= 0.00316\ h^{3} {\rm Mpc^{-3}}$. As can be seen, the distribution of the H$\alpha$ selected sample is in very good agreement with that of the SFR selection, as expected. In contrast, the distribution for the [OII]-selected sample is shifted to lower instantaneous SFR and cold gas metallicity. Still, an important fraction of [OII] emitters overlap with the SFR and H$\alpha$ selections. %{\bf Can't we quantify directly the exact numbers of overlap between the differently selected samples?}
Table~\ref{tab: overlap-fractions} shows the fraction of overlap between the ELG and SFR selections. Note that ELG and SFR selected samples have less overlap at lower number densities.

\begin{table}
    \centering
    \begin{tabular}{l|ccc}
        \hline
         $n/h^{3}\rm Mpc^{-3}$ &$\rm H\alpha$ & $\rm [OIII]$ & $\rm [OII]$ \\
        \hline
        $\rm 0.001$ & $0.81$ & $0.57$ & $0.39$  \\
        $\rm 0.00316$ & $0.91$ & $0.71$ & $0.47$  \\
        $\rm 0.01$ & $0.96$ & $0.91$ & $0.79$  \\
        \hline
    \end{tabular}
    \caption{Fraction of SFR-selected galaxies that also satisfy the ELG selection criteria. Different rows indicate the number density of the samples as shown in the first column. Columns 2, 3 and 4 give the  fraction of objects that also meet the H$\alpha$, [OIII] and [OII] selections, respectively.}
    \label{tab: overlap-fractions}
\end{table}

%%%%%%%%%%%%%%%%%%%%%%%%%%%%%%%%%%%%%%%%%%%%%%%%%%%%%%%%%%%%%%%%%%%%%%%%%%%%%%%%%%%
\section{The galaxy assembly bias of ELG samples}

Measurements from N-body simulations have shown that in order to fully determine the clustering of dark matter haloes one needs, in addition to their masses, knowledge of secondary halo properties such as formation time, concentration, subhalo occupation and spin \citep[e.g][]{Gao05, Wechsler06,Gao07}. This effect, termed halo assembly bias, may potentially have an impact on the galaxy content of haloes, producing variations in the halo occupation functions and therefore affecting the large-scale galaxy clustering amplitude \citep[e.g][]{Artale18, Zehavi18, contreras19}. Hence, it is important to model this effect when interpreting the correlation function using the standard HOD framework.

SAM samples that are obtained using halo merger histories extracted from N-body simulations are affected by assembly bias because the growth histories of dark matter haloes, and therefore the level of assembly bias that haloes are subject to, also affect the galaxies that evolve within them. To measure the impact of assembly bias on the clustering of our galaxy samples we compare their 2PCFs with that of ``shuffled'' galaxy samples. The shuffling removes information about the assembly history of haloes by randomly exchanging the galaxy populations between haloes of the same mass \citep{Croton07, Xu20b}. The standard approach preserves the central-satellites distances, therefore the one-halo terms of the shuffled catalogues are the same as those of the original SAM samples. The assembly bias signature can be obtained by comparing the 2PCF of the SAM samples to that of the shuffled samples. 

\begin{figure}
    \centering
    \includegraphics[width=\columnwidth]{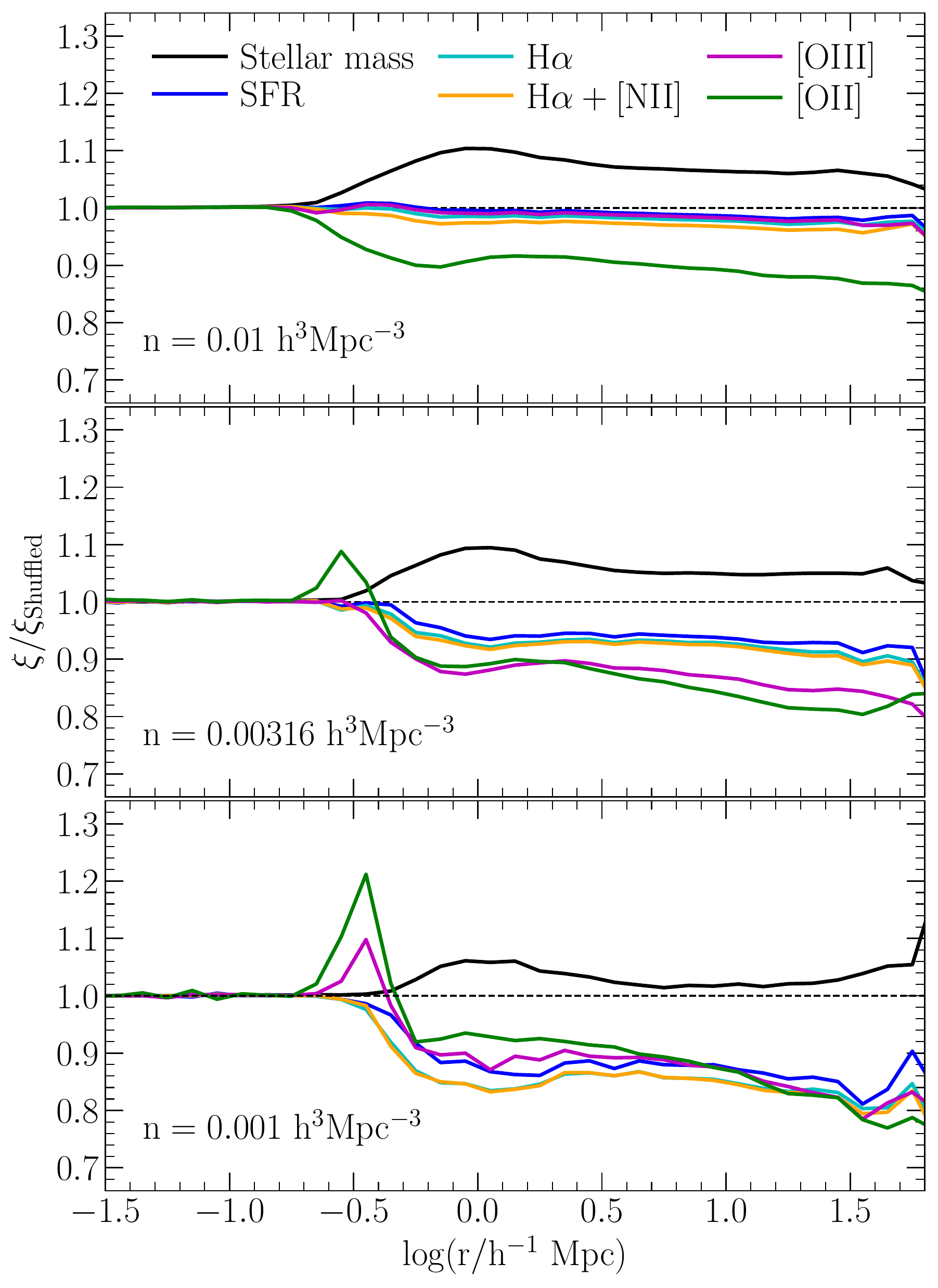}
    \caption{The assembly bias signature in the SAG samples. Each panel shows a different number density as labelled. Note that the assembly bias for the [OIII] and [OII] selections exhibits a clear scale-dependent signature for the two lowest number densities.}
    \label{fig: GAB}
\end{figure}

The impact of assembly bias on galaxy clustering depends on the selection criteria, number density and redshift of the sample \citep[e.g][]{contreras19}. As the ELG selection shows substantial overlap with selection by SFR (see Fig.~\ref{fig: corner-plot}), we can estimate how much of the effect of assembly bias on the clustering of ELGs comes from the SFR selection. We do this by looking at the assembly bias effect present in a purely SFR-selected sample. Even though nebular emission traces SFR, some properties of the gas in the ISM, such as metallicity, can introduce additional effects that are not included when selecting by SFR alone. The assembly bias signatures for SFR-, stellar mass- and ELG-selected samples are shown in Fig.~\ref{fig: GAB} for the three number densities. It can be seen from Fig.~\ref{fig: GAB} that the assembly bias in the H$\alpha$ and H$\alpha$ + [NII] selections is similar to that seen in the SFR selection. In contrast, assembly bias suppresses the galaxy clustering of [OIII] and [OII] selections by up to $30$ per cent. Table~\ref{tab: overlap-fractions} shows that the H$\alpha$ sample has a high overlap with the SFR selected sample for all number densities considered. For [OIII] emitters, the overlap with the SFR-selection is high for the highest number density sample, explaining their similar clustering in the top panel of Fig.~\ref{fig:GAB}. For the other number densities considered, the overlap between the [OIII] and SFR selected samples is much smaller and their clustering is different.

%{\bf OLD COMMENT: What's going on with the [OIII] selection (magenta line) in the top panel??}
%{\color{red} \bf EJ: The 91 per cent of galaxies in the OIII sample are contained in the SFR sample. I think this similarity between the two samples explains their almost identical 2PCFs}.{\color{green} IZ: OK!}

Furthermore, the assembly bias is scale-dependent, particularly in the case of [OII]. This steepness in the ratio of the 2PCF to the shuffled samples is also present in the SFR, H$\alpha$  and H$\alpha$+[NII] selections but only for the lowest number densities and, in any case, it is not as scale-dependent as in the [OIII] and [OII] cases. Moreover, there is a ``bump" feature in the ratio that is present only for the latter two selections around the transition from the one-halo to the two-halo term ($\log(r/h^{-1}{\rm Mpc}) \sim -0.4$). Overall, there is a clear trend between number density and the impact of assembly bias on galaxy clustering; for all ELG selections, the suppression of the clustering amplitude is larger for samples with lower number densities (i.e. for galaxies with higher emission line luminosities). Also, note that the steepness of the assembly bias signature in the [OIII] and [OII] selections is more pronounced for lower number densities. We also show the impact in the stellar mass-selected samples; it can be seen that assembly bias enhances the galaxy clustering and shows no scale-dependence for any of the number densities explored here. 

To investigate the origin of the steepness in the assembly bias signature, we analyse the galaxy clustering of the SAG and shuffled samples separately, by comparing their 2PCFs to that of the dark matter. The dark matter 2PCF is obtained from the linear power spectra used in the MDPL2 simulation, which is Fourier transformed to obtain the linear theory matter correlation function. %by inputing the cosmological parameters used in the MDPL2 simulation into the {\tt CAMB} code \citep{Lewis:2000}, which provides the matter power spectrum in linear perturbation theory\footnote{{\color{red} Note that {\tt CAMB} was used to generate the linear power spectra for the MDPL2 simulation}}. The power spectrum is then  Fourier transformed to obtain the linear theory matter correlation function.
Using the 2PCF of the dark matter, $\xi_{\rm mm}$, and the 2PCF of the galaxy sample $\xi_{\rm gal}$ we compute the large scale bias of each sample using 
\begin{equation} \label{eq: bias}
b(r) = \sqrt{\frac{\xi_{\rm gal}}{\xi_{\rm mm}}}.
\end{equation}
The value of the bias  parameter is expected to be constant to  first-order on linear scales and to vary with the galaxy selection. 

\begin{figure*}
  \begin{minipage}{\textwidth}
  \centering
      \includegraphics[width=0.98\textwidth]{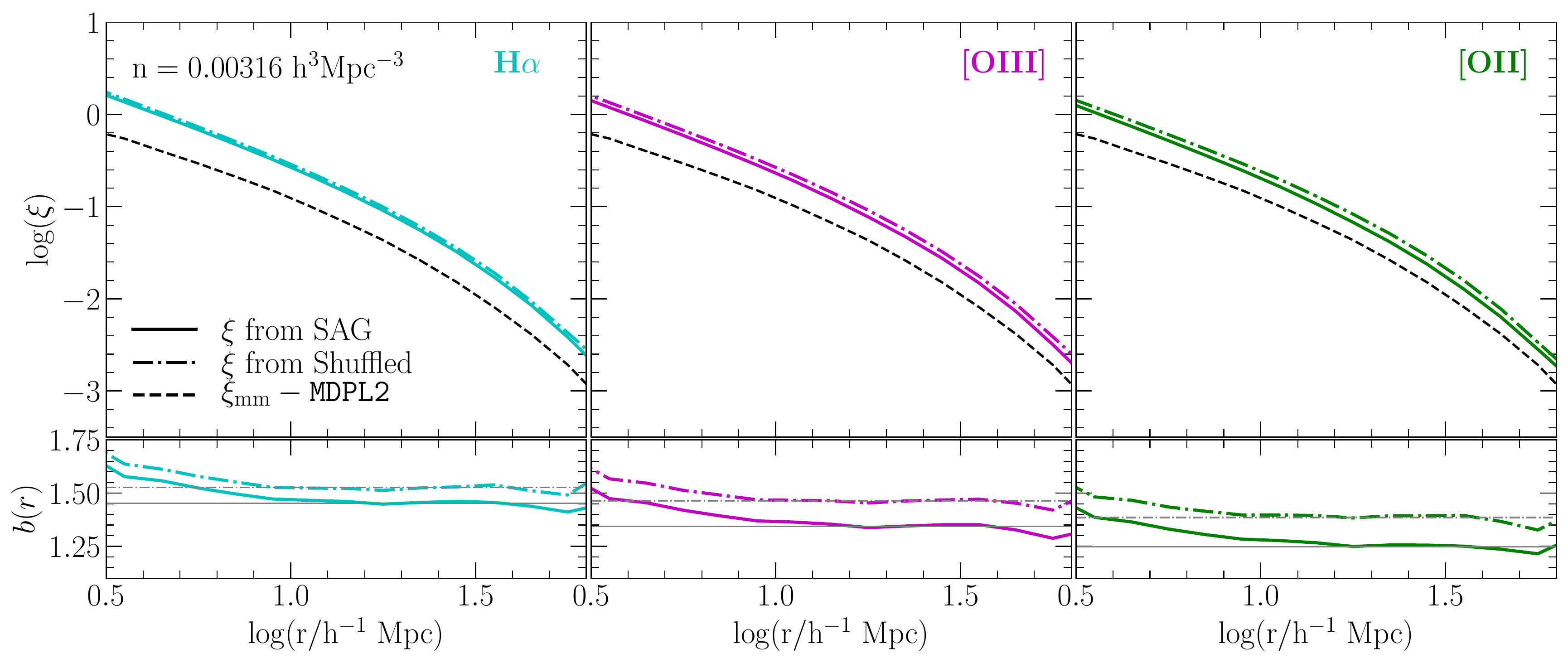}
      \caption{({\it Top}) The 2PCFs of the SAG samples (solid) and the shuffled samples (dashed-dotted) for the H$\alpha$ ({\it left panel}), [OIII] ({\it middle panel}) and [OII] ({\it right panel}) selections. The black dashed lines correspond to the 2PCF of the dark matter distribution of the MDPL2 simulation. ({\it Bottom}) The bias parameter, as a function of scale-separation, for the SAG and shuffled samples. The horizontal solid (dashed) grey line corresponds to the bias parameters of the SAG (shuffled) samples averaged between $25$ and $50\ h^{-1} \rm Mpc$.} 
     \label{fig: bias}
  \end{minipage}
\end{figure*}

The main panels of Fig.~\ref{fig: bias} show the 2PCFs of the dark matter and the SAG and  shuffled samples for the H$\alpha$, [OIII] and [OII] selections, for the number density of  $n=0.00316\ h^3 {\rm Mpc^{-3}}$. For clarity, we show results for the two-halo term only ($\log(r /h^{-1}{\rm Mpc}) \gtrsim 0.5$). The estimate of the bias parameter,  $b(r)$, for each sample is shown as a coloured line in the bottom panels. We average these values between $25$ and ${\rm 50\ }h^{-1}{\rm Mpc}$ to obtain a constant large-scale bias,  which is shown by the gray horizontal lines for comparison with $b(r)$.  For the H$\alpha$ selection the bias parameter is roughly constant over the range  ${\rm 25 < r/}h^{-1}{\rm Mpc\  < 50}$, for both the SAG and shuffled samples. In contrast, the bias parameters for the SAG [OIII] and [OII] selections show a scale-dependence.  The bias for the shuffled samples is seen to be roughly constant.  For the lowest number density sample (not shown here), we find that the bias parameter has an even steeper scale dependence for [OII] and [OIII] selections. The larger value found for the bias of the H$\alpha$ selected sample indicates that, in this case, galaxies trace higher peaks in the density field than galaxies in the other  ELG-selected samples.

The prediction of a scale-dependent bias parameter for the SAG [OIII]- and [OII]-selected samples indicates that there are additional features which shape the large scale clustering of these tracers. This suggests that the gas metallicity, which has an impact on the [OII] and [OIII] emission for a given amount of star formation, has some dependence on environment. This is confirmed by the much weaker scale dependence found for the bias on large scales in the shuffled samples. 

%%%%%%%%%%%%%%%%%%%%%%%%%%%%%%%%%%%%%%%%%%%%%%%%%%%%%%%%%%%%%%%%%%%%%%%%%%%%%%%%%%

\section{Origin of scale-dependent assembly bias}

In this section we investigate the origin of the scale-dependence of the galaxy assembly bias signature in galaxies selected by their [OIII] and [OII] line emission. As just noted, the scale-dependent bias is only present in the original SAG samples and not in their shuffled counterparts. This suggests that the preference for the environment that characterises galaxies with strong line emission, that could cause this scale dependence, is removed when shuffling these samples. \citet{GP20} analyzed how model ELGs trace the large scale structure in an N-Body simulation. They found that about half of [OII] emitters live in filaments while one-third live in sheets. This indicates that the galaxies selected using [OII] line luminosity will be preferentially located in low density regions. Hence, quantifying the effect that the shuffling procedure has of moving ELGs to random locations in the cosmic web can provide an insight into the relation between the environment of [OII] emitters and the scale-dependent assembly bias. Whilst there is some overlap between the SFR and [OII] selected samples, the SFR-selected sample shows little sign of scale dependent bias. Hence, the source of the scale dependence is likely to be found in the galaxies that are not in common between the two samples. Indeed, Table~\ref{tab: overlap-fractions} shows that the overlap between the SFR and [OII] emitters is less than 50 per cent for the two lowest density samples. This indicates that the origin of the scale dependence is encoded in the selection by [OII] luminosity.

%{\color{blue}This paragraph needs tightening up: now that you have the tables showing overlaps between samples, you can break this down a bit more -- for the highest density sample, there is strong overlap between SFR and [OIII] emitters, and 70 per cent overlap for [OII] -- for other densities there is only partial overlap 50 per cent or less for [OII]} {\color{green} agree}

One approach to quantifying the environment of a galaxy sample is to compute the local number density of main haloes around each galaxy. We use the main halos as we associate galaxies with their host dark matter haloes rather than with subhaloes; not all of the satellite galaxies may be associated with a resolved subhalo.  We define the number density, $n_{\rm local}$, using the distance to the fifth nearest main halo in the MDPL2 simulation, $r_{5}$, as  $n_{\rm local}=5/V(r_{5})$, where $ V(r_{5})$ is the volume of a sphere of radius $r=r_{5}$.  To count neighbouring haloes we use those with masses $M_{\rm h} > 10^{10.8} h^{-1} \rm M_{\odot}$ for all galaxy selections.  

\begin{figure}
    \centering
    \includegraphics[width=\columnwidth]{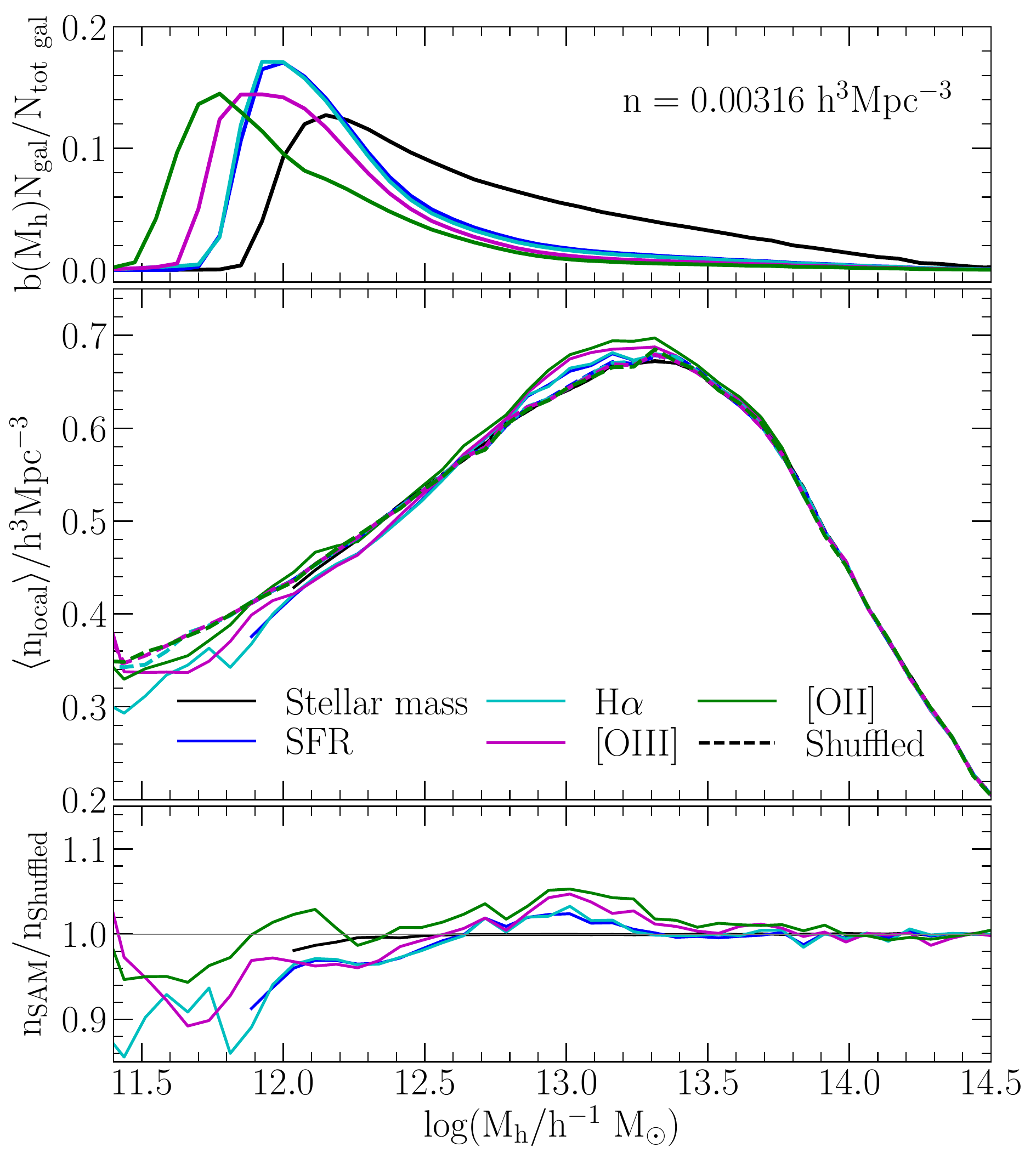}
    \caption{({\it Top}) The contribution to the effective clustering bias from halos as a function of halo mass. ({\it Middle}) the average local number densities for the SAG samples (solid) and their shuffled samples (dashed) as a function of halo mass, defined as described in the text. ({\it Bottom}) The ratio between the SAG and shuffled density measurements. Different colours indicate different galaxy selections as labelled in the middle panel.}
    \label{fig: local_den}
\end{figure}

We now consider the contribution of different halos to the sample bias and their environment. Following \citet{Kim09}, we compute the effective clustering bias as a function of halo mass for each galaxy sample, and show the results in the top panel of Fig~\ref{fig: local_den}. This parameter quantifies the contribution of galaxies in haloes of a given mass to the large scale galaxy clustering amplitude of the sample. The effective bias is simply computed as $b(M)\times \Phi(M) \times \left< N(M) \right>$ where $b(M)$ is the bias function, $ \Phi(M)$ is the halo mass function and $\left< N(M) \right>$ is the halo occupation function of the galaxy sample. For each selection, the effective bias reaches a peak at different halo masses, close to the location of the ``knee" of the occupation function (i.e. when the highest fraction of haloes of a given mass contain a central). The middle panel shows the average $n_{\rm local}$ for the SAG and shuffled samples; the bottom panel shows the ratios between the $n_{\rm local}$ of the SAG and shuffled samples for each galaxy selection. Note that a ratio higher than unity indicates that galaxies in the SAG samples are in higher density regions than their shuffled counterparts.% \st{The results in the middle and bottom panels are shown for halo masses with an effective bias larger than zero.} 

%\textbf{?? larger than unity? I thought the halo bias never got close to zero?} {\color{red} \bf EJ: That's true, but this is the {\emph{effective}} bias which has a maximum value of one. Also, note that the effective bias is the bias weighted by the HOD (and the halo mass function) which can be close to zero}
%{\color{blue}I don't get the answer -- why does the effective bias have a maximum of one? The effective bias is just a weighted average of the bias for each mass of halo -- this can go up to numbers much larger than one if the halos are massive enough, and the lowest value should be close to the lowest value of the halo bias, which is just below one. I'm missing something here.} {\color{green} I don't understand this either. Unless there's anything we need to fix here, maybe simply remove the last sentence?}

The SAG galaxy samples exhibit different average $n_{\rm local}$, which suggest that these different galaxy populations live in different environments. In contrast, the averaged $n_{\rm local}$ for the shuffled samples are largely the same for all selections. Therefore, it appears that the shuffling procedure removes correlations between the selection and environment. For example, for the [OII]-selected galaxies from SAG, the shuffling removes the strong environmental preference for [OII] emitters to reside in filaments and sheets. As the shuffling procedure moves galaxies between haloes of the same mass, the resulting HODs of the shuffled samples are similar to those of their SAG counterparts, which in turn are notably different between selections (see Fig.~\ref{fig: HODs}). Thus, their $b(M)$ and 2PFCs are also different even when their averaged $n_{\rm local}$ are similar.  
%Note that, the average mass of the host haloes of each shuffled sample is not the same, 
%\textbf{Why not? Because the shuffling bins have a finite width and the number density of halos changes across the bin? Or is the shuffling down on a bigger catalogue and then the number density cut is reapplied after the shuffling?? We might need to add an  explanation in a footnote.} {\color{red} \bf EJ: The HODs of the shuffled samples are (almost) identical to those of their SAG counterparts. Hence, after the shuffling the average mass of host haloes does not change (but the original environments does!)}
%{\color{blue} Sorry - another answer I dont understand -- why is the shuffled sample HOD different?} {\color{green} I don't understand this either; also the rest of the sentence about b(M).}
%which explains why their bias functions $b(M)$ are also different and why the shuffled samples have different large scale clustering even when their averaged $n_{\rm local}$ are similar.
For the stellar mass-selected sample, we see that the ratio of $n_{\rm local}$ in the SAG model to that in the shuffled counterpart is close to unity which indicates that the shuffling procedure does not modify the environment for this particular selection. For the SFR- and ELG-selected samples, on the other hand, the ratios in the bottom panel of Fig.~\ref{fig: local_den} show clear departures from unity. Moreover, these ratios depend on halo mass, which suggests that the mass of the host haloes is related to the strength of the environmental selection of SAG galaxies. Even though all selections show this dependence on mass, in principle, its impact on galaxy clustering only appears to be important when this dependence is seen at halo masses around the peak of the contribution to the effective clustering bias for each selection. In particular,  for the [OII] selection, there is an overlap between the peak of the effective bias and the mass dependence of the density ratio in the halo mass range $11.6 < \log(M_{\rm h}/ h^{-1}{\rm  M_{\odot}}) < 12 $. In contrast, for the SFR and H$\alpha$ selections, we find no dependence on halo mass around the peak of the effective bias in $\log(M_{\rm h}/h^{-1} {\rm M_{\odot}}) \sim 12$. For the [OIII]-selected sample, the ratio also depends on halo mass but in a narrower range ${\rm 11.7 < \log(M_h/}h^{-1}{\rm M_{\odot}}) < 11.9$, which is close to the peak of the effective bias.

\begin{figure*}
    \centering
    \begin{tabular}{c c}
         \includegraphics[width=0.48\textwidth]{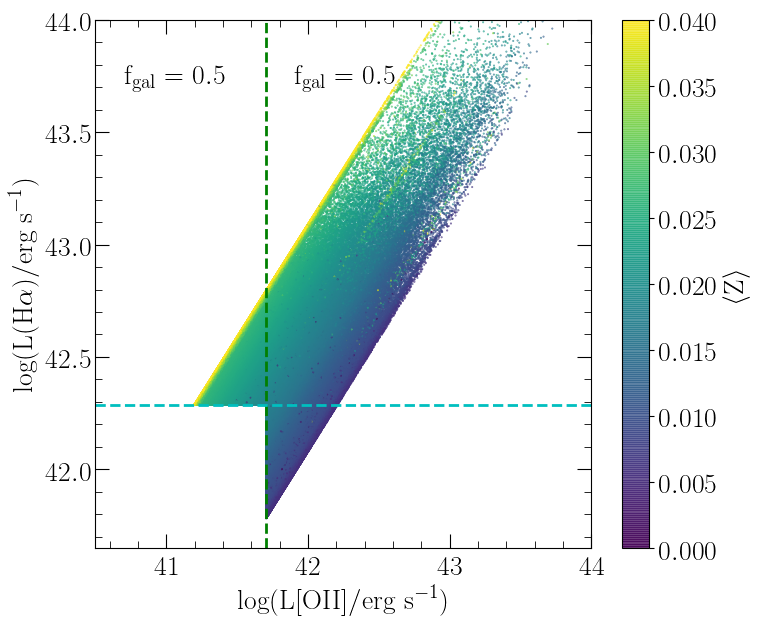}& 
         \includegraphics[width=0.48\textwidth]{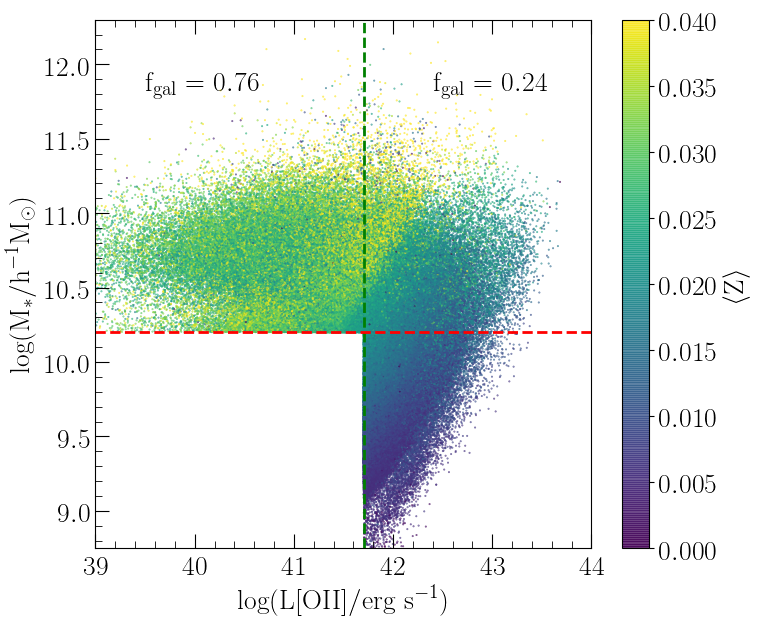} 
    \end{tabular}
    \caption{({\it Left}) The H$\alpha$ emission as a function of [OII] emission colour-coded by the cold mass-weighted metallicity of the disks and bulges. The cyan (green) dashed line indicates the cut in H$\alpha$ ([OII]) for a sample of number density $n=0.00316 \ h^{3}{\rm Mpc^{-3}}$. The fraction of galaxies with [OII] emission below and above the [OII] cut are included in both sectors. ({\it Right}) Same as the left but for stellar mass as a function of [OII] emission.} 
    \label{fig: scatter-plot-Z}
\end{figure*}

Another approach to explore the origin of the scale-dependent assembly bias signature is to analyse the distribution of the gas-phase metallicity in the SAG samples. As the {\tt GET\_EMLINES} code uses this property as an input to predict the [OIII] and [OII] emission line luminosities, we expect that it is correlated to some extent with the spatial distribution of the [OIII] and [OII] selections. Fig.~\ref{fig: scatter-plot-Z} shows the distributions of SAG galaxies in the stellar mass-L[OII] and L(H$\alpha$)-L[OII] planes. The points are colour-coded by metallicity averaged weighting by the mass of cold gas in both disks and bulges. We divide the galaxies into four subsamples separated by cuts in stellar mass, L(H$\alpha$) and L[OII] (dashed lines) that correspond to a number density $n=0.00316\ h^3 {\rm Mpc^{-3}}$. Galaxies to the right (above) of the vertical (horizontal) dashed lines are contained in the [OII]-selected sample (stellar mass and H$\alpha$ selections). In this way, the overlap between the galaxy samples can be easily seen; for the L(H$\alpha$)-L[OII] plane we see that about half of [OII]-selected galaxies are contained in the H$\alpha$ selection, while for the stellar mass-L[OII] comparison, the [OII] emitters account for 25 per cent of galaxies in the stellar mass selection. Moreover, galaxies in the [OII] selection tend to be more metal-poor than for their H$\alpha$ or stellar mass counterparts, which is consistent with the metallicity distributions in Fig.~\ref{fig: corner-plot}. Indeed, it is clear that a large fraction of the [OII]-selected galaxies, in the bottom-right sectors, are the most metal poor.

%\textbf{Just the disk? Does this include the reservoir of gas used in bursts as well? Should we plot an average of these two reservoirs for each galaxy, weighted by the mass of cold gas? Check the zgas limit of the grid in getemlines -- the higher metallicities might be saturated}. {\color{red} \bf EJ: The luminosities shown in the paper were computed using the contribution from both burst and quiescent SFR from disks. Actually, the model used by getemlines consider zgas between to 0.001 < z < 0.04.}

\begin{figure*}
    \begin{minipage}{\textwidth}
    \centering
        \includegraphics[width=\columnwidth]{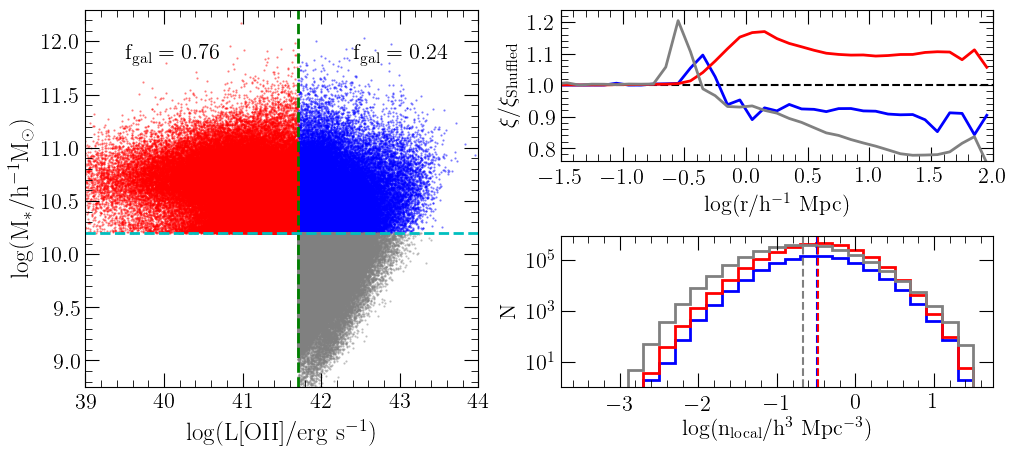}
        \caption{({\it Left}) Same as the right panel of Fig.~\ref{fig: scatter-plot-Z} but with galaxies colour-coded by their location in the stellar mass-L[OII] plane. ({\it Top-right}) the assembly bias signatures for the three colour-coded subsamples. ({\it Bottom-right}) The local density distribution for galaxies in the color-coded subsamples, where dashed vertical lines correspond to the mean of each distribution.}
        \label{fig: scatter-plot-Z-div}
    \end{minipage}
\end{figure*}

We compute the auto-correlation functions of galaxies in each sector of the stellar mass-L[OII] plane to look for interesting features in their spatial clustering. We apply the shuffling procedure to these subsamples and measure the 2PCF of the resulting shuffled samples. We also compute the ratios of the 2PCF measured from the SAG  subsamples to those of their shuffled counterparts to obtain the assembly bias signatures. The top right panel in Fig.~\ref{fig: scatter-plot-Z-div} shows the assembly bias signatures for each subsample, colour-coded by its location in the stellar mass-L[OII] plane, as indicated in the left panel. There is a remarkable difference between the assembly bias signatures of the H$\alpha$ subsamples; the red-coded galaxies, which are not included in the [OII] selection, show almost no assembly bias, but the blue-coded ones show a prominent scale-dependence. Moreover, the assembly bias for the grey-coded subsample (which mostly contains metal-poor galaxies), exhibits a steeper dependence on separation. These two results suggest that galaxies with low gas-phase metallicity are driving the scale-dependent assembly bias. To connect this information with the environment of host haloes, we show, in the bottom-right panel of Fig.~\ref{fig: scatter-plot-Z-div}, the distribution of local number densities for galaxies in each subsample. There is a subtle preference for grey-coded galaxies to  live in less dense regions than galaxies in the other subsamples. These results suggest that the gas-phase metallicity is the property of the [OIII] and [OII] selections that produces the scale-dependent assembly bias. Specifically, galaxies with low metallicity, that live in low-density regions, appear to account for most of the scale-dependence. Nevetheless, further studies are needed to fully understand the correlation between metallicity and the spatial distribution of galaxies.

%%%%%%%%%%%%%%%%%%%%%%%%%%%%%%%%%%%%%%%%%%%%%%%%%%%%%%%%%%%%%%%%%%%%%%%%%%
\section{Impact on cosmology}

In the previous sections we analyzed the scale-dependent assembly bias in the [OIII] and [OII] selected samples and its relation to the gas-phase metallicity and the environment of the galaxies. The scale dependence was found to be driven by low metallicity galaxies in underdense environments. The scale dependence may have important implications for cosmological analyses.

In this section we focus on the baryon acoustic oscillation (BAO) feature and the $\beta$ parameter.  These quantities are measured from each of the SAG samples to check if the ELG  selection introduces any systematic effects into the inference of cosmological parameters. This analysis is particularly important for the [OIII] and [OII] selections, as they contain particular features such as a non constant bias and a galaxy assembly bias signal driven by the environment of these galaxies.   

\begin{figure*}
  \begin{minipage}{\textwidth}
  \centering
  	\includegraphics[width=0.98\textwidth]{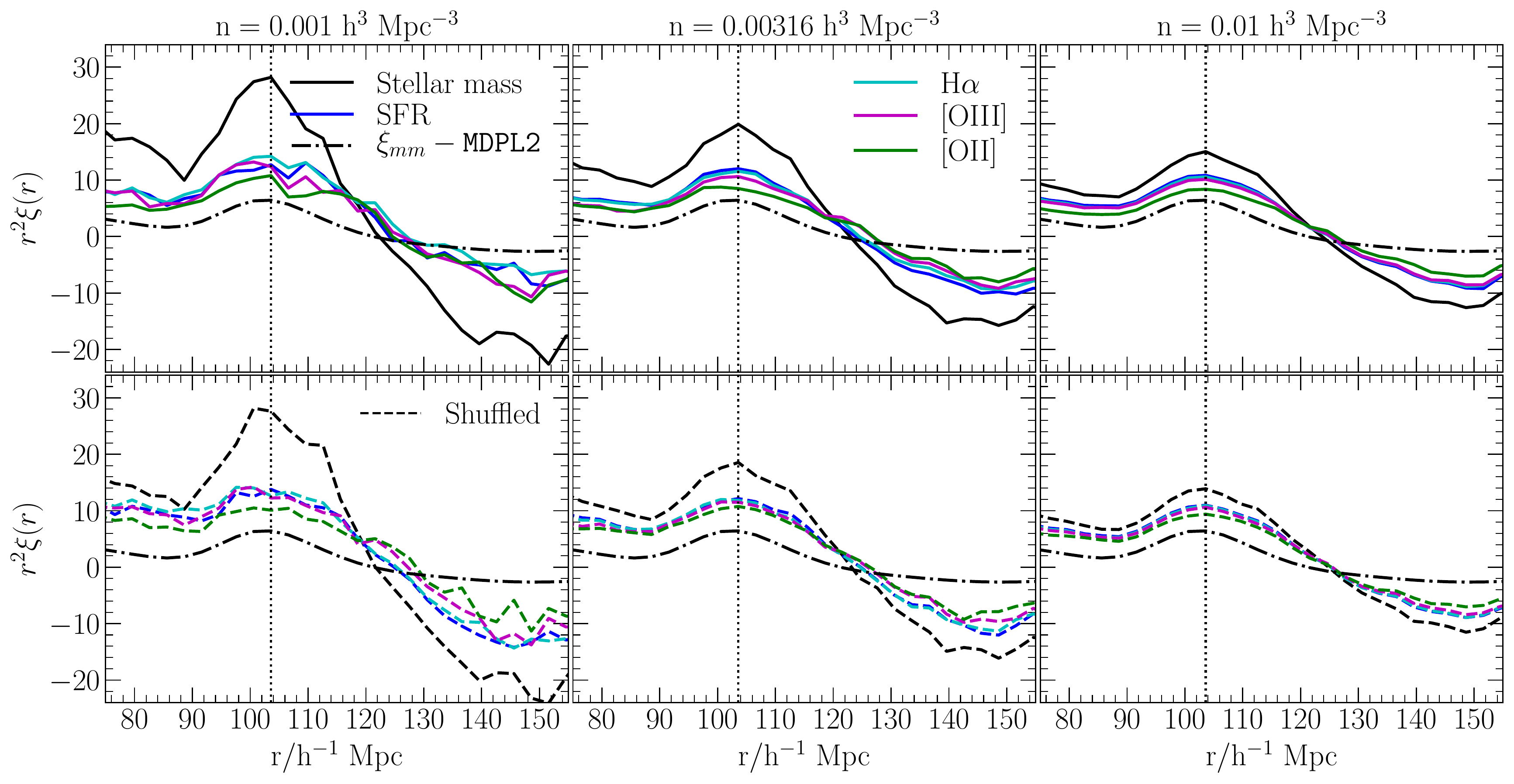}
  	\caption{The baryon acoustic oscillation (BAO) feature for the SAG samples ({\it top}) and their shuffled counterparts ({\it bottom}) with number densities $n= 0.001\ h^{3}{\rm Mpc^{-3}}$ ({\it left}, $n= 0.00316\ h^{3}{\rm Mpc^{-3}}$ ({\it middle}) and $ n= 0.00316\ h^{3}{\rm Mpc^{-3}}$ ({\it right}). Colours correspond to the different selections adopted, as shown in the keys. The vertical lines are included to guide the eye and indicate the position of the BAO peak for the dark matter, according to the linear theory prediction (black dot-dashed line).}
 	\label{fig: BAO}
  \end{minipage}
\end{figure*}

Fig.~\ref{fig: BAO} shows the 2PCFs of the SAG (top panels) and shuffled samples (bottom panels) for different selection criteria and number densities.  Note that, in order to focus on the BAO peak, we display $r^2 \times \xi(r)$. For comparison, we show the $\rm z \sim 1$ linear theory prediction for the dark matter 2PCF of the MDPL2. The vertical dotted lines mark the position of the BAO peak for the dark matter.

The amplitudes of the BAO peaks are affected by the large-scale bias of each sample, which depends on the selection criteria, number density and the degree of assembly bias present in the samples (recall the wide range of values found for the large-scale bias for the SAG and shuffled samples in Fig~\ref{fig: bias}). Interestingly, the shape of the BAO peak for the SAG [OII]-selected sample with the intermediate number density (top-middle panel), is notably different with respect to the other samples. The BAO peak position for this selection differs at the $\sim 3$ per cent level from the scale predicted by linear perturbation theory. For the lowest number density, the [OII] selection does not show a clear shift in the BAO scale, although it must be noted that all 2PCFs are somewhat noisy due to the sparsity of this number density. For the highest number density, the BAO scale for the [OII] selection is in agreement with the prediction from linear theory. This result suggests that selecting galaxies with lower [OII] line luminosity washes out the correlation between the [OII] selection and environment (see \S~5). Moreover, note that the BAO peak in the [OII] shuffled sample (bottom-middle panel) is not shifted as in the SAG counterpart which suggests that the shuffling procedure removes the shift in the BAO peak. Nevertheless, it is still unclear whether the $\sim 3$ per cent shift would result in a biased measurement of cosmological parameters as the environment effects which influence the clustering measurements for the [OII] sample could also be interpreted as a  non-local bias, which can be introduced as a nuisance parameter in BAO peak analyses (e.g. \citealt{Sanchez:2008}).

We also compute the $\beta$ parameter for the SAG samples. This parameter is a function of the logarithmic growth rate, which depends on the matter density parameter, and the bias parameter of the galaxy sample. Following \citet{Padilla19}, we compute $\beta$ for shuffled samples where the relative velocities of the galaxies within haloes -- in addition to their positions -- are maintained when shuffling satellites between haloes. The $\beta$ parameter can be obtained from the ratio between the monopoles of the correlation functions in real and redshift space \citep{Kaiser86}:

\begin{equation} \label{eq: beta_parameter}
	\xi_0(s) = \left( 1 + \frac{2}{3}\beta + \frac{1}{5}\beta^2 \right)\xi(r).
\end{equation}{}

\begin{figure}
	\centering
	\includegraphics[width=\columnwidth]{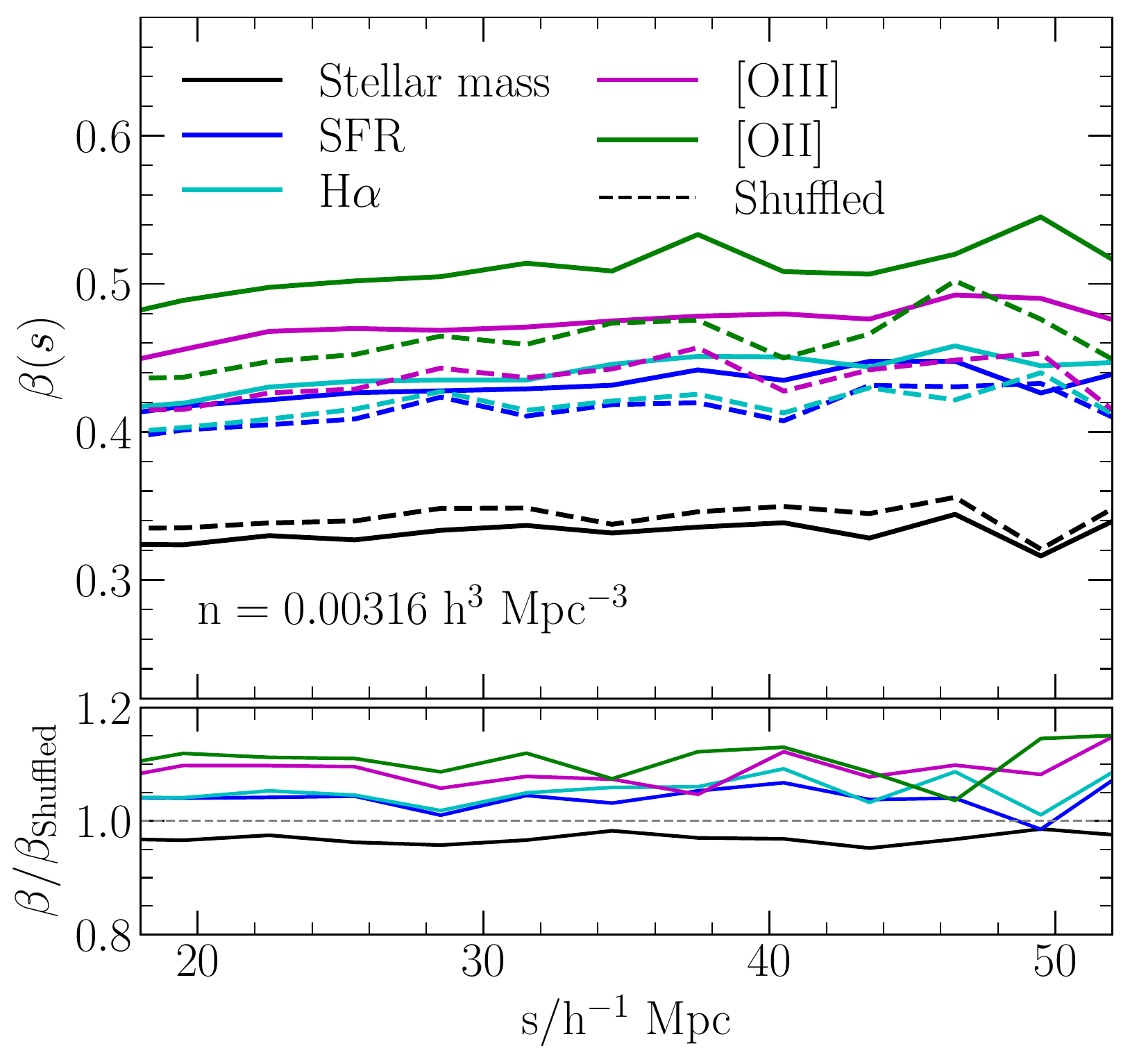}
	\caption{({\it Top}) The $\beta$ parameter as a function of separation for the SAG (solid) and shuffled samples (dotted) with number density $n= 0.00316\ h^{3} {\rm Mpc^{-3}}$. Different colours correspond to different selection criteria as indicated by the keys. ({\it Bottom}) the ratio between the $\beta$ parameters of the SAG and shuffled samples. The gray dashed line indicates unit ratio. }
	\label{fig: beta}
\end{figure}

The main panel of Fig.~\ref{fig: beta} shows the $\beta$ parameter, calculated from Eq.~\ref{eq: beta_parameter}, as a function of scale for the SAG (solid) and shuffled samples (dashed) with a number density of $n= 0.00316\ h^{3}{\rm Mpc^{-3}}$. For the SAG samples, we see that the $\beta$ parameter is roughly constant for the stellar mass selection, and it is scale-dependent for the other selections. However, for the shuffled samples, the scale dependence of $\beta$ remains.
The bottom panel of Fig.~\ref{fig: beta} shows the ratios between the $\beta$ parameter of the SAG samples to that of their shuffled counterparts. For the stellar mass selection, the ratio is almost constant for all scales. For SFR- and ELG-selected samples, the ratio is more affected by noise, but it appears to be consistent with a constant value, with the SFR and ELG selected samples returning a higher value of beta than their shuffled counterparts. In principle, the steepness of $\beta$ could be so slight that it falls below our noise level, even though we do find a scale-dependent bias for the ELG SAG samples, and a roughly constant one for the shuffled catalogues, especially for the [OIII] and [OII] selections (see Fig.~\ref{fig: bias}).

%%%%%%%%%%%%%%%%%%%%%%%%%%%%%%%%%%%%%%%%%%%%%%%%%%%%%%%%
\section{Conclusions}

The next generation of galaxy surveys such as DESI and Euclid will map the sky by measuring redshifts to unprecedented numbers of emission-line galaxies. To fully exploit this upcoming data we need to understand how these galaxies trace the underlying density of the Universe and to establish if there are any systematic effects which might impair our ability to extract unbiased cosmological information.

To investigate the potential of ELGs to constrain cosmological parameters, we study the clustering and halo occupation of the model galaxies from the SAG semi-analytical model \citep{Cora18} applied to the MDPL2 simulation outputs \citep{Klypin16}. We use the instantaneous SFR and gas-phase metallicities of galaxies as the inputs to the {\tt GET\_EMLINES} code to obtain the nebular line emission of the galaxies. To mimic the selection criteria of future surveys, we use fixed number density samples where galaxies are ranked according to their [OIII], [OII], H$\alpha$ and H$\alpha$+[NII] line luminosities. For comparison, we also include SFR- and stellar mass-selected samples.

We measure the two-point correlation functions (2PCFs), the halo occupation distributions (HODs) and the galaxy assembly bias signatures for each galaxy sample.  The galaxy assembly bias is measured via the ratio between the 2PCF of a SAG sample with that of a shuffled version of the sample, which, by construction, does not contain assembly bias. We also compute the absolute large scale bias of ELGs to look for correlations between assembly bias, large-scale bias, the environment of the galaxies, and the gas-phase metallicity of the ELG-selected samples. Finally, we measure the BAO feature and the $\beta$ parameter for the SAG and shuffled samples to investigate the implications for cosmological studies using ELGs. Our results can be summarised as follows:

\begin{itemize}
    \item ELG-selected samples have 2PCFs and HODs that are similar to those of SFR selected galaxies. However, the [OIII]- and [OII]-selected samples are less clustered than either the SFR or H$\alpha$ samples. Moreover, their HODs indicate that most of the selected galaxies live in low mass haloes. These differences explain why selecting by the luminosity of the [OIII] or [OII] lines does not reproduce the behaviour of a SFR selected sample.
    
    \item The assembly bias signature (i.e. the ratio between the 2PCFs measured for the SAG and shuffled samples) for the [OIII] and [OII] selected samples is scale-dependent, with a steepness which becomes more pronounced for lower density samples (higher [OIII] and [OII] thresholds). For the SFR and H$\alpha$ selections, the assembly bias is scale-dependent for samples with the lowest number density. In the case of galaxies selected by stellar mass, the assembly bias is roughly constant for all number densities.
    
    \item The large-scale bias, defined as the ratio between the 2PCF of a galaxy sample and that of the dark matter, is scale-dependent for the [OIII]- and [OII]-selected samples in the SAG model. For the shuffled samples, in contrast, the large-scale bias is roughly constant for all selections. This suggests that the shuffling procedure removes an encoded dependence between the galaxy properties of [OIII] and [OII] selections with the environment.
    %{\bf ???? (i.e. is the galaxy assembly bias the one that has an scale dependent signal, not the regular bias of the galaxy sample) \color{red} EJ: Both the regular bias (b(r)) and the galaxy assembly bias are scale-dependent for[OIII] and [OII] selections}.
    
    \item For a fixed halo mass the local number density - that we use to quantify the environment of host haloes - is roughly the same for all shuffled samples. This indicates that the shuffling procedure eliminates the correlation between the selection and the environment of host haloes. In contrast, for the SAG samples the local number densities are notably different between the selection criteria which indicates that in some cases the selected galaxies live in special regions of the cosmic web. Moreover, the change of the environment of the [OII]-selected sample has a strong dependence with halo mass (see Fig.~\ref{fig: local_den}).
    
    \item Galaxies with low gas-phase metallicities are the ones that produce the scale-dependent assembly bias signature. Indeed, for the SAG sample, a larger fraction of metal-poor galaxies results in a steeper scale-dependent assembly bias signature  (see Fig.~\ref{fig: scatter-plot-Z-div}). Moreover, these galaxies tend to live in low-density regions, which is more common for the ELG selected samples.
    
    \item The scale of the BAO feature in the [OII]-selected sample with number density $n= 0.00316\ h^{3}{\rm Mpc^{-3}}$, differs at the 3 per cent level from that recovered using the other selections. Interestingly, this shift in the BAO peak is not seen for the shuffled counterpart of the [OII] sample. Conversely, for the other selected samples, the BAO feature is located at the same position for both the SAG and  shuffled samples. There is no clear shift in the position of the BAO feature for the [OII] selections with lower and higher number densities.
    
    \item The $\beta$ parameter for the SFR- and ELG-selected samples is non-constant as a function of scale for the SAG and the shuffled samples. This is clearer for the [OII] and [OIII] selections and can be explained as a combination of the scale-dependent large scale bias and a possible non-constant logarithmic growth rate. For the stellar mass case, in contrast, $\beta$ is roughly constant.
\end{itemize}

Our results show that care must be given when using future galaxy samples from Euclid and DESI, which will be selected by their emission line luminosities.  We find that this type of selection can produce samples that lie in special environments, and because of this their clustering can show a different slope than that of the underlying matter density field.  This type of environment selection needs to be modelled and marginalised over in cosmological parameter constraints from such samples in order to avoid systematic effects in their analysis.

\section*{Acknowledgements}

This work was made possible by the efforts of Gerard Lemson and colleagues at the German Astronomical Virtual Observatory in setting up the Millennium Simulation database in Garching. The CosmoSim database used in this paper is a service by the Leibniz-Institute for Astrophysics Potsdam (AIP). The MultiDark database was developed in cooperation with the Spanish MultiDark Consolider Project CSD2009-00064. The author gratefully acknowledge the Gauss Centre for Supercomputing e.V. (www.gauss-centre.eu) and the Partnership for Advanced Supercomputing in Europe (PRACE, www.prace-ri.eu) for funding the MultiDark simulation project by providing computing time on the GCS Supercomputer SuperMUC at Leibniz Supercomputing Centre (LRZ, www.lrz.de).
EJ  acknowledges  support  from  ``Centro  de  Astronom\'{i}a  y  Tecnolog\'{i}as  Afines'' BASAL 170002. IZ acknowledges support by NSF grant AST-1612085. This  project  has  received  funding  from  the European  Union's  Horizon  2020  Research  and  Innovation Programme under the Marie Sk\l{}odowska-Curie grant agreement No 734374. The calculations for this paper were performed  on  the  Geryon  computer  at  the  Center  for Astro-Engineering UC, part of the BASAL PFB-06, which received additional  funding  from  QUIMAL 130008  and  Fondequip AIC-57 for upgrades.

\section*{Data availability statement}

The data underlying this article will be shared on reasonable request to the corresponding author. 

%%%%%%%%%%%%%%%%%%%%%%%%%%%%%%%%%%%%%%%%%%%%%%%%%%

%%%%%%%%%%%%%%%%%%%% REFERENCES %%%%%%%%%%%%%%%%%%

\bibliography{Biblio} 

\begin{thebibliography}{}
\makeatletter
\relax
\def\mn@urlcharsother{\let\do\@makeother \do\$\do\&\do\#\do\^\do\_\do\%\do\~}
\def\mn@doi{\begingroup\mn@urlcharsother \@ifnextchar [ {\mn@doi@}
  {\mn@doi@[]}}
\def\mn@doi@[#1]#2{\def\@tempa{#1}\ifx\@tempa\@empty \href
  {http://dx.doi.org/#2} {doi:#2}\else \href {http://dx.doi.org/#2} {#1}\fi
  \endgroup}
\def\mn@eprint#1#2{\mn@eprint@#1:#2::\@nil}
\def\mn@eprint@arXiv#1{\href {http://arxiv.org/abs/#1} {{\tt arXiv:#1}}}
\def\mn@eprint@dblp#1{\href {http://dblp.uni-trier.de/rec/bibtex/#1.xml}
  {dblp:#1}}
\def\mn@eprint@#1:#2:#3:#4\@nil{\def\@tempa {#1}\def\@tempb {#2}\def\@tempc
  {#3}\ifx \@tempc \@empty \let \@tempc \@tempb \let \@tempb \@tempa \fi \ifx
  \@tempb \@empty \def\@tempb {arXiv}\fi \@ifundefined
  {mn@eprint@\@tempb}{\@tempb:\@tempc}{\expandafter \expandafter \csname
  mn@eprint@\@tempb\endcsname \expandafter{\@tempc}}}

\bibitem[\protect\citeauthoryear{{Artale}, {Zehavi}, {Contreras}  \&
  {Norberg}}{{Artale} et~al.}{2018}]{Artale18}
{Artale} M.~C.,  {Zehavi} I.,  {Contreras} S.,   {Norberg} P.,  2018, \mn@doi
  [MNRAS] {10.1093/mnras/sty2110}, \href
  {https://ui.adsabs.harvard.edu/abs/2018MNRAS.480.3978A} {480, 3978}

\bibitem[\protect\citeauthoryear{{Baldwin}, {Phillips}  \&
  {Terlevich}}{{Baldwin} et~al.}{1981}]{BPT:1981}
{Baldwin} J.~A.,  {Phillips} M.~M.,   {Terlevich} R.,  1981, \mn@doi [\pasp]
  {10.1086/130766}, \href
  {https://ui.adsabs.harvard.edu/abs/1981PASP...93....5B} {93, 5}

\bibitem[\protect\citeauthoryear{{Baugh}}{{Baugh}}{2006}]{Baugh06-rv}
{Baugh} C.~M.,  2006, \mn@doi [Reports on Progress in Physics]
  {10.1088/0034-4885/69/12/R02}, \href
  {http://adsabs.harvard.edu/abs/2006RPPh...69.3101B} {69, 3101}

\bibitem[\protect\citeauthoryear{{Baugh} et~al.,}{{Baugh}
  et~al.}{2019}]{Baugh19}
{Baugh} C.~M.,  et~al., 2019, \mn@doi [\mnras] {10.1093/mnras/sty3427}, \href
  {https://ui.adsabs.harvard.edu/abs/2019MNRAS.483.4922B} {483, 4922}

\bibitem[\protect\citeauthoryear{{Bautista} et~al.,}{{Bautista}
  et~al.}{2018}]{Bautista18}
{Bautista} J.~E.,  et~al., 2018, \mn@doi [\apj] {10.3847/1538-4357/aacea5},
  \href {https://ui.adsabs.harvard.edu/abs/2018ApJ...863..110B} {863, 110}

\bibitem[\protect\citeauthoryear{{Behroozi}, {Wechsler}  \& {Wu}}{{Behroozi}
  et~al.}{2013a}]{Behroozi2013a}
{Behroozi} P.~S.,  {Wechsler} R.~H.,   {Wu} H.-Y.,  2013a, \mn@doi [ApJ]
  {10.1088/0004-637X/762/2/109}, \href
  {https://ui.adsabs.harvard.edu/abs/2013ApJ...762..109B} {762, 109}

\bibitem[\protect\citeauthoryear{{Behroozi}, {Wechsler}, {Wu}, {Busha},
  {Klypin}  \& {Primack}}{{Behroozi} et~al.}{2013b}]{Behroozi2013b}
{Behroozi} P.~S.,  {Wechsler} R.~H.,  {Wu} H.-Y.,  {Busha} M.~T.,  {Klypin}
  A.~A.,   {Primack} J.~R.,  2013b, \mn@doi [ApJ] {10.1088/0004-637X/763/1/18},
  \href {https://ui.adsabs.harvard.edu/abs/2013ApJ...763...18B} {763, 18}

\bibitem[\protect\citeauthoryear{{Benson}}{{Benson}}{2012}]{Benson12}
{Benson} A.~J.,  2012, \mn@doi [\na] {10.1016/j.newast.2011.07.004}, \href
  {https://ui.adsabs.harvard.edu/abs/2012NewA...17..175B} {17, 175}

\bibitem[\protect\citeauthoryear{{Benson}, {Cole}, {Frenk}, {Baugh}  \&
  {Lacey}}{{Benson} et~al.}{2000}]{Benson00}
{Benson} A.~J.,  {Cole} S.,  {Frenk} C.~S.,  {Baugh} C.~M.,   {Lacey} C.~G.,
  2000, \mn@doi [MNRAS] {10.1046/j.1365-8711.2000.03101.x}, \href
  {http://adsabs.harvard.edu/abs/2000MNRAS.311..793B} {311, 793}

\bibitem[\protect\citeauthoryear{{Berlind} \& {Weinberg}}{{Berlind} \&
  {Weinberg}}{2002}]{Berlind02}
{Berlind} A.~A.,  {Weinberg} D.~H.,  2002, \mn@doi [ApJ] {10.1086/341469},
  \href {http://adsabs.harvard.edu/abs/2002ApJ...575..587B} {575, 587}

\bibitem[\protect\citeauthoryear{{Byler}, {Dalcanton}, {Conroy}  \&
  {Johnson}}{{Byler} et~al.}{2017}]{Byler:2017}
{Byler} N.,  {Dalcanton} J.~J.,  {Conroy} C.,   {Johnson} B.~D.,  2017, \mn@doi
  [\apj] {10.3847/1538-4357/aa6c66}, \href
  {https://ui.adsabs.harvard.edu/abs/2017ApJ...840...44B} {840, 44}

\bibitem[\protect\citeauthoryear{{Cole}, {Lacey}, {Baugh}  \& {Frenk}}{{Cole}
  et~al.}{2000}]{cole2000}
{Cole} S.,  {Lacey} C.~G.,  {Baugh} C.~M.,   {Frenk} C.~S.,  2000, \mn@doi
  [MNRAS] {10.1046/j.1365-8711.2000.03879.x}, \href
  {http://adsabs.harvard.edu/abs/2000MNRAS.319..168C} {319, 168}

\bibitem[\protect\citeauthoryear{{Cole} et~al.,}{{Cole} et~al.}{2005}]{Cole05}
{Cole} S.,  et~al., 2005, \mn@doi [\mnras] {10.1111/j.1365-2966.2005.09318.x},
  \href {https://ui.adsabs.harvard.edu/abs/2005MNRAS.362..505C} {362, 505}

\bibitem[\protect\citeauthoryear{{Contreras}, {Baugh}, {Norberg}  \&
  {Padilla}}{{Contreras} et~al.}{2013}]{contreras13}
{Contreras} S.,  {Baugh} C.~M.,  {Norberg} P.,   {Padilla} N.,  2013, \mn@doi
  [MNRAS] {10.1093/mnras/stt629}, \href
  {http://adsabs.harvard.edu/abs/2013MNRAS.432.2717C} {432, 2717}

\bibitem[\protect\citeauthoryear{{Contreras}, {Zehavi}, {Padilla}, {Baugh},
  {Jim{\'e}nez}  \& {Lacerna}}{{Contreras} et~al.}{2019}]{contreras19}
{Contreras} S.,  {Zehavi} I.,  {Padilla} N.,  {Baugh} C.~M.,  {Jim{\'e}nez} E.,
    {Lacerna} I.,  2019, \mn@doi [MNRAS] {10.1093/mnras/stz018}, \href
  {http://adsabs.harvard.edu/abs/2019MNRAS.484.1133C} {484, 1133}

\bibitem[\protect\citeauthoryear{{Contreras}, {Angulo}  \&
  {Zennaro}}{{Contreras} et~al.}{2020}]{Contreras20}
{Contreras} S.,  {Angulo} R.,   {Zennaro} M.,  2020, arXiv e-prints, \href
  {https://ui.adsabs.harvard.edu/abs/2020arXiv200503672C} {p. arXiv:2005.03672}

\bibitem[\protect\citeauthoryear{{Cora}}{{Cora}}{2006}]{Cora06}
{Cora} S.~A.,  2006, \mn@doi [MNRAS] {10.1111/j.1365-2966.2006.10271.x}, \href
  {https://ui.adsabs.harvard.edu/abs/2006MNRAS.368.1540C} {368, 1540}

\bibitem[\protect\citeauthoryear{{Cora} et~al.,}{{Cora} et~al.}{2018}]{Cora18}
{Cora} S.~A.,  et~al., 2018, \mn@doi [\mnras] {10.1093/mnras/sty1131}, \href
  {https://ui.adsabs.harvard.edu/abs/2018MNRAS.479....2C} {479, 2}

\bibitem[\protect\citeauthoryear{{Croton} et~al.,}{{Croton}
  et~al.}{2006}]{Croton06}
{Croton} D.~J.,  et~al., 2006, \mn@doi [MNRAS]
  {10.1111/j.1365-2966.2005.09675.x}, \href
  {http://adsabs.harvard.edu/abs/2006MNRAS.365...11C} {365, 11}

\bibitem[\protect\citeauthoryear{{Croton}, {Gao}  \& {White}}{{Croton}
  et~al.}{2007}]{Croton07}
{Croton} D.~J.,  {Gao} L.,   {White} S.~D.~M.,  2007, \mn@doi [MNRAS]
  {10.1111/j.1365-2966.2006.11230.x}, \href
  {http://adsabs.harvard.edu/abs/2007MNRAS.374.1303C} {374, 1303}

\bibitem[\protect\citeauthoryear{{Croton} et~al.,}{{Croton}
  et~al.}{2016}]{Croton16}
{Croton} D.~J.,  et~al., 2016, \mn@doi [\apjs] {10.3847/0067-0049/222/2/22},
  \href {https://ui.adsabs.harvard.edu/abs/2016ApJS..222...22C} {222, 22}

\bibitem[\protect\citeauthoryear{{DESI Collaboration} et~al.,}{{DESI
  Collaboration} et~al.}{2016}]{DESI16}
{DESI Collaboration} et~al., 2016, arXiv e-prints, \href
  {https://ui.adsabs.harvard.edu/abs/2016arXiv161100036D} {p. arXiv:1611.00036}

\bibitem[\protect\citeauthoryear{{Dawson} et~al.,}{{Dawson}
  et~al.}{2013}]{Dawson13}
{Dawson} K.~S.,  et~al., 2013, \mn@doi [\aj] {10.1088/0004-6256/145/1/10},
  \href {https://ui.adsabs.harvard.edu/abs/2013AJ....145...10D} {145, 10}

\bibitem[\protect\citeauthoryear{{Dawson} et~al.,}{{Dawson}
  et~al.}{2016}]{Dawson16}
{Dawson} K.~S.,  et~al., 2016, \mn@doi [\aj] {10.3847/0004-6256/151/2/44},
  \href {https://ui.adsabs.harvard.edu/abs/2016AJ....151...44D} {151, 44}

\bibitem[\protect\citeauthoryear{{De Lucia} \& {Blaizot}}{{De Lucia} \&
  {Blaizot}}{2007}]{DeLucia07}
{De Lucia} G.,  {Blaizot} J.,  2007, \mn@doi [MNRAS]
  {10.1111/j.1365-2966.2006.11287.x}, \href
  {http://adsabs.harvard.edu/abs/2007MNRAS.375....2D} {375, 2}

\bibitem[\protect\citeauthoryear{{De Lucia}, {Kauffmann}  \& {White}}{{De
  Lucia} et~al.}{2004}]{DeLucia04}
{De Lucia} G.,  {Kauffmann} G.,   {White} S.~D.~M.,  2004, \mn@doi [MNRAS]
  {10.1111/j.1365-2966.2004.07584.x}, \href
  {http://adsabs.harvard.edu/abs/2004MNRAS.349.1101D} {349, 1101}

\bibitem[\protect\citeauthoryear{{Dopita} \& {Sutherland}}{{Dopita} \&
  {Sutherland}}{1995}]{Dopita95}
{Dopita} M.~A.,  {Sutherland} R.~S.,  1995, \mn@doi [ApJ] {10.1086/176596},
  \href {https://ui.adsabs.harvard.edu/abs/1995ApJ...455..468D} {455, 468}

\bibitem[\protect\citeauthoryear{{Eisenstein} et~al.,}{{Eisenstein}
  et~al.}{2005}]{Eisenstein05}
{Eisenstein} D.~J.,  et~al., 2005, \mn@doi [\apj] {10.1086/466512}, \href
  {https://ui.adsabs.harvard.edu/abs/2005ApJ...633..560E} {633, 560}

\bibitem[\protect\citeauthoryear{{Eisenstein} et~al.,}{{Eisenstein}
  et~al.}{2011}]{Eisenstein11}
{Eisenstein} D.~J.,  et~al., 2011, \mn@doi [\aj] {10.1088/0004-6256/142/3/72},
  \href {https://ui.adsabs.harvard.edu/abs/2011AJ....142...72E} {142, 72}

\bibitem[\protect\citeauthoryear{{Favole} et~al.,}{{Favole}
  et~al.}{2016}]{Favole16}
{Favole} G.,  et~al., 2016, \mn@doi [\mnras] {10.1093/mnras/stw1483}, \href
  {https://ui.adsabs.harvard.edu/abs/2016MNRAS.461.3421F} {461, 3421}

\bibitem[\protect\citeauthoryear{{Favole} et~al.,}{{Favole}
  et~al.}{2020}]{Favole20}
{Favole} G.,  et~al., 2020, \mn@doi [\mnras] {10.1093/mnras/staa2292}, \href
  {https://ui.adsabs.harvard.edu/abs/2020MNRAS.tmp.2333F} {}

\bibitem[\protect\citeauthoryear{{Gao} \& {White}}{{Gao} \&
  {White}}{2007}]{Gao07}
{Gao} L.,  {White} S. D.~M.,  2007, \mn@doi [\mnras]
  {10.1111/j.1745-3933.2007.00292.x}, \href
  {https://ui.adsabs.harvard.edu/abs/2007MNRAS.377L...5G} {377, L5}

\bibitem[\protect\citeauthoryear{{Gao}, {Springel}  \& {White}}{{Gao}
  et~al.}{2005}]{Gao05}
{Gao} L.,  {Springel} V.,   {White} S.~D.~M.,  2005, \mn@doi [MNRAS]
  {10.1111/j.1745-3933.2005.00084.x}, \href
  {http://adsabs.harvard.edu/abs/2005MNRAS.363L..66G} {363, L66}

\bibitem[\protect\citeauthoryear{{Gargiulo} et~al.,}{{Gargiulo}
  et~al.}{2015}]{Gargiulo2015}
{Gargiulo} I.~D.,  et~al., 2015, \mn@doi [MNRAS] {10.1093/mnras/stu2272}, \href
  {https://ui.adsabs.harvard.edu/abs/2015MNRAS.446.3820G} {446, 3820}

\bibitem[\protect\citeauthoryear{{Gonzalez-Perez} et~al.,}{{Gonzalez-Perez}
  et~al.}{2018}]{gp18}
{Gonzalez-Perez} V.,  et~al., 2018, \mn@doi [MNRAS] {10.1093/mnras/stx2807},
  \href {http://adsabs.harvard.edu/abs/2018MNRAS.474.4024G} {474, 4024}

\bibitem[\protect\citeauthoryear{{Gonzalez-Perez} et~al.,}{{Gonzalez-Perez}
  et~al.}{2020}]{GP20}
{Gonzalez-Perez} V.,  et~al., 2020, arXiv e-prints, \href
  {https://ui.adsabs.harvard.edu/abs/2020arXiv200106560G} {p. arXiv:2001.06560}

\bibitem[\protect\citeauthoryear{{Groves} \& {Allen}}{{Groves} \&
  {Allen}}{2010}]{Groves10}
{Groves} B.~A.,  {Allen} M.~G.,  2010, \mn@doi [\na]
  {10.1016/j.newast.2010.02.005}, \href
  {https://ui.adsabs.harvard.edu/abs/2010NewA...15..614G} {15, 614}

\bibitem[\protect\citeauthoryear{{Groves}, {Dopita}  \& {Sutherland}}{{Groves}
  et~al.}{2004}]{Groves04}
{Groves} B.~A.,  {Dopita} M.~A.,   {Sutherland} R.~S.,  2004, \mn@doi [ApJS]
  {10.1086/421113}, \href
  {https://ui.adsabs.harvard.edu/abs/2004ApJS..153....9G} {153, 9}

\bibitem[\protect\citeauthoryear{{Guo} et~al.,}{{Guo} et~al.}{2011}]{guo11}
{Guo} Q.,  et~al., 2011, \mn@doi [MNRAS] {10.1111/j.1365-2966.2010.18114.x},
  \href {http://adsabs.harvard.edu/abs/2011MNRAS.413..101G} {413, 101}

\bibitem[\protect\citeauthoryear{{Guo}, {White}, {Angulo}, {Henriques},
  {Lemson}, {Boylan-Kolchin}, {Thomas}  \& {Short}}{{Guo} et~al.}{2013}]{Guo13}
{Guo} Q.,  {White} S.,  {Angulo} R.~E.,  {Henriques} B.,  {Lemson} G.,
  {Boylan-Kolchin} M.,  {Thomas} P.,   {Short} C.,  2013, \mn@doi [MNRAS]
  {10.1093/mnras/sts115}, \href
  {http://adsabs.harvard.edu/abs/2013MNRAS.428.1351G} {428, 1351}

\bibitem[\protect\citeauthoryear{{Gutkin}, {Charlot}  \& {Bruzual}}{{Gutkin}
  et~al.}{2016}]{Gutkin:2016}
{Gutkin} J.,  {Charlot} S.,   {Bruzual} G.,  2016, \mn@doi [\mnras]
  {10.1093/mnras/stw1716}, \href
  {https://ui.adsabs.harvard.edu/abs/2016MNRAS.462.1757G} {462, 1757}

\bibitem[\protect\citeauthoryear{{Henriques}, {White}, {Thomas}, {Angulo},
  {Guo}, {Lemson}  \& {Springel}}{{Henriques} et~al.}{2013}]{Henriques13}
{Henriques} B.~M.~B.,  {White} S.~D.~M.,  {Thomas} P.~A.,  {Angulo} R.~E.,
  {Guo} Q.,  {Lemson} G.,   {Springel} V.,  2013, \mn@doi [MNRAS]
  {10.1093/mnras/stt415}, \href
  {http://adsabs.harvard.edu/abs/2013MNRAS.431.3373H} {431, 3373}

\bibitem[\protect\citeauthoryear{{Henriques}, {White}, {Thomas}, {Angulo},
  {Guo}, {Lemson}, {Springel}  \& {Overzier}}{{Henriques}
  et~al.}{2015}]{Henriques15}
{Henriques} B.~M.~B.,  {White} S.~D.~M.,  {Thomas} P.~A.,  {Angulo} R.,  {Guo}
  Q.,  {Lemson} G.,  {Springel} V.,   {Overzier} R.,  2015, \mn@doi [MNRAS]
  {10.1093/mnras/stv705}, \href
  {http://adsabs.harvard.edu/abs/2015MNRAS.451.2663H} {451, 2663}

\bibitem[\protect\citeauthoryear{{Jim{\'e}nez}, {Contreras}, {Padilla},
  {Zehavi}, {Baugh}  \& {Gonzalez-Perez}}{{Jim{\'e}nez}
  et~al.}{2019}]{Jimenez19}
{Jim{\'e}nez} E.,  {Contreras} S.,  {Padilla} N.,  {Zehavi} I.,  {Baugh} C.~M.,
    {Gonzalez-Perez} V.,  2019, \mn@doi [\mnras] {10.1093/mnras/stz2790}, \href
  {https://ui.adsabs.harvard.edu/abs/2019MNRAS.490.3532J} {490, 3532}

\bibitem[\protect\citeauthoryear{{Kaiser}}{{Kaiser}}{1986a}]{Kaiser86a}
{Kaiser} N.,  1986a, \mn@doi [\mnras] {10.1093/mnras/219.4.785}, \href
  {https://ui.adsabs.harvard.edu/abs/1986MNRAS.219..785K} {219, 785}

\bibitem[\protect\citeauthoryear{{Kaiser}}{{Kaiser}}{1986b}]{Kaiser86}
{Kaiser} N.,  1986b, \mn@doi [MNRAS] {10.1093/mnras/222.2.323}, \href
  {https://ui.adsabs.harvard.edu/abs/1986MNRAS.222..323K} {222, 323}

\bibitem[\protect\citeauthoryear{{Kim}, {Baugh}, {Cole}, {Frenk}  \&
  {Benson}}{{Kim} et~al.}{2009}]{Kim09}
{Kim} H.-S.,  {Baugh} C.~M.,  {Cole} S.,  {Frenk} C.~S.,   {Benson} A.~J.,
  2009, \mn@doi [\mnras] {10.1111/j.1365-2966.2009.15560.x}, \href
  {https://ui.adsabs.harvard.edu/abs/2009MNRAS.400.1527K} {400, 1527}

\bibitem[\protect\citeauthoryear{{Klypin}, {Yepes}, {Gottl{\"o}ber}, {Prada}
  \& {He{\ss}}}{{Klypin} et~al.}{2016}]{Klypin16}
{Klypin} A.,  {Yepes} G.,  {Gottl{\"o}ber} S.,  {Prada} F.,   {He{\ss}} S.,
  2016, \mn@doi [MNRAS] {10.1093/mnras/stw248}, \href
  {https://ui.adsabs.harvard.edu/abs/2016MNRAS.457.4340K} {457, 4340}

\bibitem[\protect\citeauthoryear{{Knebe} et~al.,}{{Knebe}
  et~al.}{2018}]{knebe18}
{Knebe} A.,  et~al., 2018, \mn@doi [\mnras] {10.1093/mnras/stx2662}, \href
  {https://ui.adsabs.harvard.edu/abs/2018MNRAS.474.5206K} {474, 5206}

\bibitem[\protect\citeauthoryear{{Kravtsov}, {Berlind}, {Wechsler}, {Klypin},
  {Gottl{\"o}ber}, {Allgood}  \& {Primack}}{{Kravtsov}
  et~al.}{2004}]{Kravtsov04}
{Kravtsov} A.~V.,  {Berlind} A.~A.,  {Wechsler} R.~H.,  {Klypin} A.~A.,
  {Gottl{\"o}ber} S.,  {Allgood} B.,   {Primack} J.~R.,  2004, \mn@doi [ApJ]
  {10.1086/420959}, \href {http://adsabs.harvard.edu/abs/2004ApJ...609...35K}
  {609, 35}

\bibitem[\protect\citeauthoryear{{Lagos}, {Cora}  \& {Padilla}}{{Lagos}
  et~al.}{2008}]{Lagos08}
{Lagos} C. D.~P.,  {Cora} S.~A.,   {Padilla} N.~D.,  2008, \mn@doi [MNRAS]
  {10.1111/j.1365-2966.2008.13456.x}, \href
  {https://ui.adsabs.harvard.edu/abs/2008MNRAS.388..587L} {388, 587}

\bibitem[\protect\citeauthoryear{{Lagos}, {Tobar}, {Robotham}, {Obreschkow},
  {Mitchell}, {Power}  \& {Elahi}}{{Lagos} et~al.}{2018}]{Lagos18}
{Lagos} C.~d.~P.,  {Tobar} R.~J.,  {Robotham} A.~S.~G.,  {Obreschkow} D.,
  {Mitchell} P.~D.,  {Power} C.,   {Elahi} P.~J.,  2018, \mn@doi [MNRAS]
  {10.1093/mnras/sty2440}, \href
  {https://ui.adsabs.harvard.edu/abs/2018MNRAS.481.3573L} {481, 3573}

\bibitem[\protect\citeauthoryear{{Laureijs} et~al.,}{{Laureijs}
  et~al.}{2011}]{Laureijs:2011}
{Laureijs} R.,  et~al., 2011, arXiv e-prints, \href
  {https://ui.adsabs.harvard.edu/abs/2011arXiv1110.3193L} {p. arXiv:1110.3193}

\bibitem[\protect\citeauthoryear{{Levesque}, {Kewley}  \& {Larson}}{{Levesque}
  et~al.}{2010}]{Levesque10}
{Levesque} E.~M.,  {Kewley} L.~J.,   {Larson} K.~L.,  2010, \mn@doi [\aj]
  {10.1088/0004-6256/139/2/712}, \href
  {https://ui.adsabs.harvard.edu/abs/2010AJ....139..712L} {139, 712}

\bibitem[\protect\citeauthoryear{{Merson}, {Wang}, {Benson}, {Faisst},
  {Masters}, {Kiessling}  \& {Rhodes}}{{Merson} et~al.}{2018}]{Merson:2018}
{Merson} A.,  {Wang} Y.,  {Benson} A.,  {Faisst} A.,  {Masters} D.,
  {Kiessling} A.,   {Rhodes} J.,  2018, \mn@doi [\mnras]
  {10.1093/mnras/stx2649}, \href
  {https://ui.adsabs.harvard.edu/abs/2018MNRAS.474..177M} {474, 177}

\bibitem[\protect\citeauthoryear{{Montero-Dorta} et~al.,}{{Montero-Dorta}
  et~al.}{2020}]{Montero20}
{Montero-Dorta} A.~D.,  et~al., 2020, \mn@doi [\mnras]
  {10.1093/mnras/staa1624}, \href
  {https://ui.adsabs.harvard.edu/abs/2020MNRAS.496.1182M} {496, 1182}

\bibitem[\protect\citeauthoryear{{Mu{\~n}oz Arancibia}, {Navarrete}, {Padilla},
  {Cora}, {Gawiser}, {Kurczynski}  \& {Ruiz}}{{Mu{\~n}oz Arancibia}
  et~al.}{2015}]{Munoz15}
{Mu{\~n}oz Arancibia} A.~M.,  {Navarrete} F.~P.,  {Padilla} N.~D.,  {Cora}
  S.~A.,  {Gawiser} E.,  {Kurczynski} P.,   {Ruiz} A.~N.,  2015, \mn@doi
  [MNRAS] {10.1093/mnras/stu2237}, \href
  {https://ui.adsabs.harvard.edu/abs/2015MNRAS.446.2291M} {446, 2291}

\bibitem[\protect\citeauthoryear{{Nagao}, {Maiolino}  \& {Marconi}}{{Nagao}
  et~al.}{2006}]{Nagao06}
{Nagao} T.,  {Maiolino} R.,   {Marconi} A.,  2006, \mn@doi [\aap]
  {10.1051/0004-6361:20065216}, \href
  {https://ui.adsabs.harvard.edu/abs/2006A&A...459...85N} {459, 85}

\bibitem[\protect\citeauthoryear{{Nelson} et~al.,}{{Nelson}
  et~al.}{2018}]{Nelson18}
{Nelson} D.,  et~al., 2018, \mn@doi [\mnras] {10.1093/mnras/stx3040}, \href
  {https://ui.adsabs.harvard.edu/abs/2018MNRAS.475..624N} {475, 624}

\bibitem[\protect\citeauthoryear{{Orsi}, {Padilla}, {Groves}, {Cora}, {Tecce},
  {Gargiulo}  \& {Ruiz}}{{Orsi} et~al.}{2014}]{Orsi14}
{Orsi} {\'A}.,  {Padilla} N.,  {Groves} B.,  {Cora} S.,  {Tecce} T.,
  {Gargiulo} I.,   {Ruiz} A.,  2014, \mn@doi [MNRAS] {10.1093/mnras/stu1203},
  \href {http://adsabs.harvard.edu/abs/2014MNRAS.443..799O} {443, 799}

\bibitem[\protect\citeauthoryear{{Padilla}, {Salazar-Albornoz}, {Contreras},
  {Cora}  \& {Ruiz}}{{Padilla} et~al.}{2014}]{Padilla14}
{Padilla} N.~D.,  {Salazar-Albornoz} S.,  {Contreras} S.,  {Cora} S.~A.,
  {Ruiz} A.~N.,  2014, \mn@doi [\mnras] {10.1093/mnras/stu1321}, \href
  {https://ui.adsabs.harvard.edu/abs/2014MNRAS.443.2801P} {443, 2801}

\bibitem[\protect\citeauthoryear{{Padilla}, {Contreras}, {Zehavi}, {Baugh}  \&
  {Norberg}}{{Padilla} et~al.}{2019}]{Padilla19}
{Padilla} N.,  {Contreras} S.,  {Zehavi} I.,  {Baugh} C.~M.,   {Norberg} P.,
  2019, \mn@doi [\mnras] {10.1093/mnras/stz824}, \href
  {https://ui.adsabs.harvard.edu/abs/2019MNRAS.486..582P} {486, 582}

\bibitem[\protect\citeauthoryear{{Planck Collaboration} et~al.,}{{Planck
  Collaboration} et~al.}{2014}]{Plank14}
{Planck Collaboration} et~al., 2014, \mn@doi [A\&A]
  {10.1051/0004-6361/201321591}, \href
  {https://ui.adsabs.harvard.edu/abs/2014A%26A...571A..16P} {571, A16}

\bibitem[\protect\citeauthoryear{{Ruiz} et~al.,}{{Ruiz} et~al.}{2015}]{Ruiz15}
{Ruiz} A.~N.,  et~al., 2015, \mn@doi [\apj] {10.1088/0004-637X/801/2/139},
  \href {https://ui.adsabs.harvard.edu/abs/2015ApJ...801..139R} {801, 139}

\bibitem[\protect\citeauthoryear{{S{\'a}nchez}, {Baugh}  \&
  {Angulo}}{{S{\'a}nchez} et~al.}{2008}]{Sanchez:2008}
{S{\'a}nchez} A.~G.,  {Baugh} C.~M.,   {Angulo} R.~E.,  2008, \mn@doi [\mnras]
  {10.1111/j.1365-2966.2008.13769.x}, \href
  {https://ui.adsabs.harvard.edu/abs/2008MNRAS.390.1470S} {390, 1470}

\bibitem[\protect\citeauthoryear{{Schaye} et~al.,}{{Schaye}
  et~al.}{2015}]{Schaye15}
{Schaye} J.,  et~al., 2015, \mn@doi [MNRAS] {10.1093/mnras/stu2058}, \href
  {http://adsabs.harvard.edu/abs/2015MNRAS.446..521S} {446, 521}

\bibitem[\protect\citeauthoryear{{Scoccimarro}, {Feldman}, {Fry}  \&
  {Frieman}}{{Scoccimarro} et~al.}{2001}]{Scoccimarro01}
{Scoccimarro} R.,  {Feldman} H.~A.,  {Fry} J.~N.,   {Frieman} J.~A.,  2001,
  \mn@doi [ApJ] {10.1086/318284}, \href
  {http://adsabs.harvard.edu/abs/2001ApJ...546..652S} {546, 652}

\bibitem[\protect\citeauthoryear{{Shim} \& {Chary}}{{Shim} \&
  {Chary}}{2013}]{Shim13}
{Shim} H.,  {Chary} R.-R.,  2013, \mn@doi [\apj] {10.1088/0004-637X/765/1/26},
  \href {https://ui.adsabs.harvard.edu/abs/2013ApJ...765...26S} {765, 26}

\bibitem[\protect\citeauthoryear{{Sinha} \& {Garrison}}{{Sinha} \&
  {Garrison}}{2017}]{Sinha17}
{Sinha} M.,  {Garrison} L.,  2017, {Corrfunc: Blazing fast correlation
  functions on the CPU}, Astrophysics Source Code Library (\mn@eprint {ascl}
  {1703.003})

\bibitem[\protect\citeauthoryear{{Somerville}, {Hopkins}, {Cox}, {Robertson}
  \& {Hernquist}}{{Somerville} et~al.}{2008}]{Somervile08}
{Somerville} R.~S.,  {Hopkins} P.~F.,  {Cox} T.~J.,  {Robertson} B.~E.,
  {Hernquist} L.,  2008, \mn@doi [\mnras] {10.1111/j.1365-2966.2008.13805.x},
  \href {https://ui.adsabs.harvard.edu/abs/2008MNRAS.391..481S} {391, 481}

\bibitem[\protect\citeauthoryear{{Springel}, {White}, {Tormen}  \&
  {Kauffmann}}{{Springel} et~al.}{2001}]{Springel01}
{Springel} V.,  {White} S.~D.~M.,  {Tormen} G.,   {Kauffmann} G.,  2001,
  \mn@doi [MNRAS] {10.1046/j.1365-8711.2001.04912.x}, \href
  {http://adsabs.harvard.edu/abs/2001MNRAS.328..726S} {328, 726}

\bibitem[\protect\citeauthoryear{{Springel} et~al.,}{{Springel}
  et~al.}{2005}]{Springel05}
{Springel} V.,  et~al., 2005, \mn@doi [Nature] {10.1038/nature03597}, \href
  {http://adsabs.harvard.edu/abs/2005Natur.435..629S} {435, 629}

\bibitem[\protect\citeauthoryear{{Stevens}, {Croton}  \& {Mutch}}{{Stevens}
  et~al.}{2016}]{Stevens16}
{Stevens} A. R.~H.,  {Croton} D.~J.,   {Mutch} S.~J.,  2016, \mn@doi [\mnras]
  {10.1093/mnras/stw1332}, \href
  {https://ui.adsabs.harvard.edu/abs/2016MNRAS.461..859S} {461, 859}

\bibitem[\protect\citeauthoryear{{Tecce}, {Cora}, {Tissera}, {Abadi}  \&
  {Lagos}}{{Tecce} et~al.}{2010}]{Tecce10}
{Tecce} T.~E.,  {Cora} S.~A.,  {Tissera} P.~B.,  {Abadi} M.~G.,   {Lagos} C.
  D.~P.,  2010, \mn@doi [MNRAS] {10.1111/j.1365-2966.2010.17262.x}, \href
  {https://ui.adsabs.harvard.edu/abs/2010MNRAS.408.2008T} {408, 2008}

\bibitem[\protect\citeauthoryear{{Vogelsberger} et~al.,}{{Vogelsberger}
  et~al.}{2014}]{Vogelsberger14}
{Vogelsberger} M.,  et~al., 2014, \mn@doi [MNRAS] {10.1093/mnras/stu1536},
  \href {http://adsabs.harvard.edu/abs/2014MNRAS.444.1518V} {444, 1518}

\bibitem[\protect\citeauthoryear{{Wechsler} \& {Tinker}}{{Wechsler} \&
  {Tinker}}{2018}]{Wechsler18}
{Wechsler} R.~H.,  {Tinker} J.~L.,  2018, \mn@doi [\araa]
  {10.1146/annurev-astro-081817-051756}, \href
  {https://ui.adsabs.harvard.edu/abs/2018ARA&A..56..435W} {56, 435}

\bibitem[\protect\citeauthoryear{{Wechsler}, {Zentner}, {Bullock}, {Kravtsov}
  \& {Allgood}}{{Wechsler} et~al.}{2006}]{Wechsler06}
{Wechsler} R.~H.,  {Zentner} A.~R.,  {Bullock} J.~S.,  {Kravtsov} A.~V.,
  {Allgood} B.,  2006, \mn@doi [ApJ] {10.1086/507120}, \href
  {http://adsabs.harvard.edu/abs/2006ApJ...652...71W} {652, 71}

\bibitem[\protect\citeauthoryear{{Weinberg}, {Mortonson}, {Eisenstein},
  {Hirata}, {Riess}  \& {Rozo}}{{Weinberg} et~al.}{2013}]{Weinberg13}
{Weinberg} D.~H.,  {Mortonson} M.~J.,  {Eisenstein} D.~J.,  {Hirata} C.,
  {Riess} A.~G.,   {Rozo} E.,  2013, \mn@doi [\physrep]
  {10.1016/j.physrep.2013.05.001}, \href
  {https://ui.adsabs.harvard.edu/abs/2013PhR...530...87W} {530, 87}

\bibitem[\protect\citeauthoryear{{Xu} \& {Zheng}}{{Xu} \&
  {Zheng}}{2020}]{Xu20a}
{Xu} X.,  {Zheng} Z.,  2020, \mn@doi [\mnras] {10.1093/mnras/staa009}, \href
  {https://ui.adsabs.harvard.edu/abs/2020MNRAS.492.2739X} {492, 2739}

\bibitem[\protect\citeauthoryear{{Xu}, {Zehavi}  \& {Contreras}}{{Xu}
  et~al.}{2020}]{Xu20b}
{Xu} X.,  {Zehavi} I.,   {Contreras} S.,  2020, arXiv e-prints, \href
  {https://ui.adsabs.harvard.edu/abs/2020arXiv200705545X} {p. arXiv:2007.05545}

\bibitem[\protect\citeauthoryear{{Zehavi} et~al.,}{{Zehavi}
  et~al.}{2011}]{Zehavi11}
{Zehavi} I.,  et~al., 2011, \mn@doi [ApJ] {10.1088/0004-637X/736/1/59}, \href
  {http://adsabs.harvard.edu/abs/2011ApJ...736...59Z} {736, 59}

\bibitem[\protect\citeauthoryear{{Zehavi}, {Contreras}, {Padilla}, {Smith},
  {Baugh}  \& {Norberg}}{{Zehavi} et~al.}{2018}]{Zehavi18}
{Zehavi} I.,  {Contreras} S.,  {Padilla} N.,  {Smith} N.~J.,  {Baugh} C.~M.,
  {Norberg} P.,  2018, \mn@doi [ApJ] {10.3847/1538-4357/aaa54a}, \href
  {http://adsabs.harvard.edu/abs/2018ApJ...853...84Z} {853, 84}

\bibitem[\protect\citeauthoryear{{Zehavi}, {Kerby}, {Contreras}, {Jim{\'e}nez},
  {Padilla}  \& {Baugh}}{{Zehavi} et~al.}{2019}]{Zehavi19}
{Zehavi} I.,  {Kerby} S.~E.,  {Contreras} S.,  {Jim{\'e}nez} E.,  {Padilla} N.,
    {Baugh} C.~M.,  2019, \mn@doi [\apj] {10.3847/1538-4357/ab4d4d}, \href
  {https://ui.adsabs.harvard.edu/abs/2019ApJ...887...17Z} {887, 17}

\bibitem[\protect\citeauthoryear{{Zheng} et~al.,}{{Zheng}
  et~al.}{2005}]{Zheng05}
{Zheng} Z.,  et~al., 2005, \mn@doi [ApJ] {10.1086/466510}, \href
  {http://adsabs.harvard.edu/abs/2005ApJ...633..791Z} {633, 791}

\makeatother
\end{thebibliography}
\bibliographystyle{mnras}

%%%%%%%%%%%%%%%%%%%%%%%%%%%%%%%%%%%%%%%%%%%%%%%%%%

%%%%%%%%%%%%%%%%% APPENDICES %%%%%%%%%%%%%%%%%%%%%

\appendix

\section{Assembly bias of ELGs in {\tt L-GALAXIES}}

To assess if the scale-dependent assembly bias can be found in other semi-analytical models, we select galaxy samples at z=0 from the \citet{Guo13} model (hereafter G13) which is a version of the {\tt L-GALAXIES} code from the Munich group \citep{DeLucia04, Croton06, DeLucia07, guo11, Henriques13}. G13 is a semi-analytic model, and as such it models a set of physical processes that shape the formation and evolution of galaxies, applied to halo merger trees drawn from the Millennium-WMAP7 simulation. This simulation was carried out in a box of 500 $h^{-1}$ Mpc a side, and is the same as the original Millennium simulation \citep{Springel05} but with updated cosmological parameters that match the results from the WMAP7 observations.

We use the {\tt GET\_EMLINES} code to obtain the nebular emission for G13 galaxies. The instantaneous SFR is not a direct output of G13, hence, we use the \emph{average} SFR instead to infer line emission luminosities. This is motivated by the results of \citet{Favole20}; they demonstrate that using average SFRs as inputs for {\tt GET\_EMLINES} produces good predictions to study average populations of [OII] emission-line galaxies. We then define new stellar mass, SFR, H$\alpha$, [OIII] and [OII] selected samples following the procedure in \S~2.4.

\begin{figure}
    \centering
    \includegraphics[width=\columnwidth]{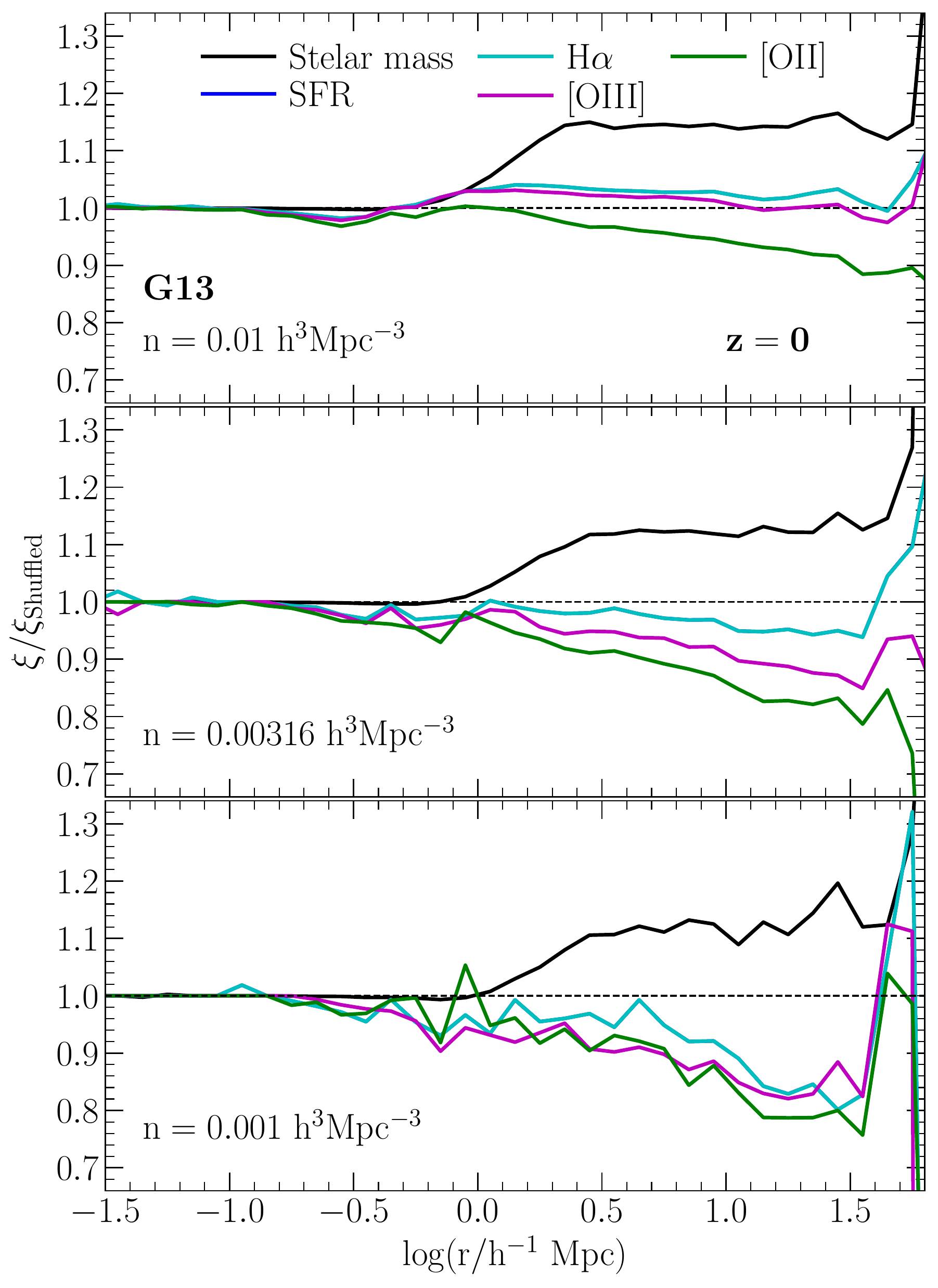}
    \caption{Same as Fig.~\ref{fig: GAB} but for galaxy samples extracted from the \citet{Guo13} SAM.}
    \label{fig: GAB of G13}
\end{figure}

Fig.~\ref{fig: GAB of G13} shows the assembly bias signatures for the G13 samples with three different number densities (the same ones used to define the SAG samples). Note that the 2PCFs of the H$\alpha$ and SFR selections are the same, which is a consequence of H$\alpha$ luminosity having a simple dependence on the SFR, with little variation with the cold gas metallicity. Even though the clustering measurements are noisier at very large scales, we still find that the [OIII] and [OII] selections have a clear scale-dependent assembly bias. In contrast, for the SFR and H$\alpha$ selections, the scale-dependence is there for the lowest number density sample alone, while for the stellar mass-selected samples the signature is roughly flat in all cases.

\section{Completeness of ELG selected samples}

Selecting by emission line luminosities produces samples that trace the amplitude of SFR but also other additional properties, such as the cold gas metallicity. Thus, in principle, a low-SFR galaxy may be included in an ELG-selected sample.
Because of this we analyze the effect of the moderate stellar mass cut imposed on the SAG data, which is present in the subsamples analyzed in this work; this moderate cut is $ M_{*} > 10^{8.7} h^{-1} {\rm M_{\odot}}$ which is slightly lower than the resolution of the MDPL2 and Millennium simulations (${\rm \sim 10^9}\ h^{-1} {\rm M_{\odot}}$).

\begin{figure}
    \centering
    \includegraphics[width=\columnwidth]{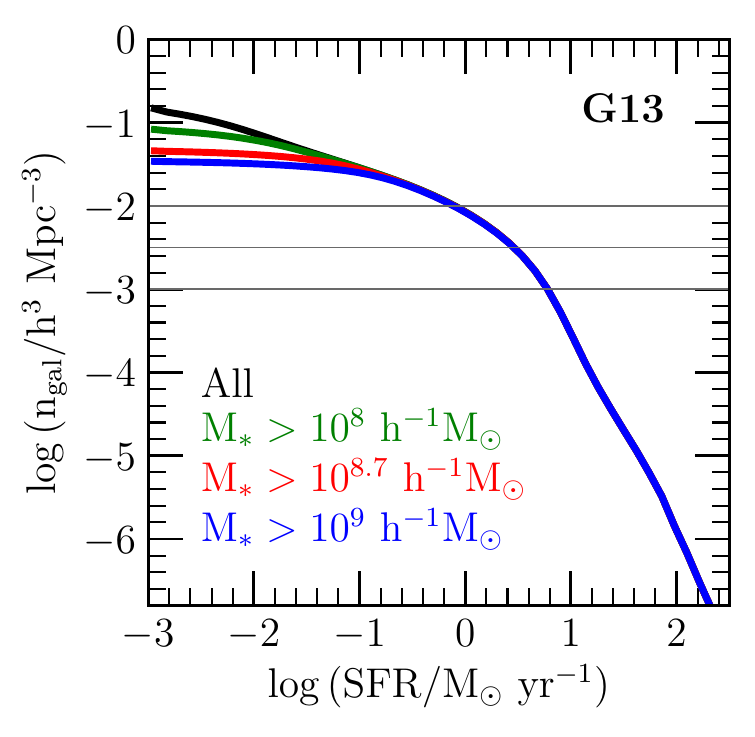}
    \caption{The cumulative SFR function of subsamples of the \citet{Guo13} model defined by different stellar mass cuts indicated by the colors and labels. Horizontal lines indicate the different number densities used to define the galaxy samples.}
    \label{fig: CumFunc}
\end{figure}

As SAG and G13 show similar trends for assembly bias (see Fig.~\ref{fig: GAB of G13}), we expect that the effect of the completeness stellar mass cut on these trends should be also similar for both models. Fig.~\ref{fig: CumFunc} displays the cumulative SFR function for subsamples of G13, defined by different stellar mass cuts. As expected from the stellar mass-SFR relation, we see that the larger the cut, the smaller the number of low-SFR galaxies.

We define a subsample of G13 by selecting galaxies with stellar mass above the cut imposed for SAG. Then, following the procedure in \S.~2.4, we define our galaxy samples with this new cut. We measure the assembly bias signatures for these samples, and we compare them with the assembly bias of G13 samples with no previous cuts in stellar mass. We find that the assembly bias signatures are almost identical for all selections and number densities. Noticeable differences only arise at very large scales(${\rm \log( r/h^{-1}\  Mpc) > 1.5}$). This indicates that the moderate stellar mass cut introduced to impose completeness has only a minor impact on the assembly bias signature, which includes the scale-dependence for [OIII] and [OII] selections. 

%%%%%%%%%%%%%%%%%%%%%%%%%%%%%%%%%%%%%%%%%%%%%%%%%%

\label{lastpage}
\end{document}